\documentclass[a4paper,11pt]{article}
\usepackage{jheppub}
\usepackage{orcidlink}
\usepackage{caption}
\usepackage{subcaption}
\usepackage{setspace}
\usepackage{hyperref}
\title{\boldmath Scalar, Vector Perturbations and Effective Hawking Radiation of Cylindrical Black Holes in $f(\mathcal{R})$ and Ricci-Inverse Gravity}

\author[a]{Faizuddin Ahmed\orcidlink{0000-0003-2196-9622},}
\author[b,1]{\.{I}zzet Sakall{\i}\orcidlink{0000-0001-7827-9476}\note{Corresponding author}}
\author[c]{Ahmad Al-Badawi\orcidlink{0000-0002-3127-3453},}
\author[d]{and Abdelmalek Bouzenada\orcidlink{0000-0002-3363-980X}}

\affiliation[a]{Department of Physics, University of Science \& Technology Meghalaya, Ri-Bhoi, Meghalaya, 793101, India}
\affiliation[c]{AS245 Department of Physics, Eastern Mediterranean University, 99628, Famagusta Northern Cyprus, via Mersin 10, Turkiye}
\affiliation[b]{Department of Physics, Al-Hussein Bin Talal University, 71111, Ma’an, Jordan}
\affiliation[d]{Laboratory of Theoretical and Applied Physics, Echahid Cheikh Larbi Tebessi University 12001, Algeria}

\emailAdd{faizuddinahmed15@gmail.com}
\emailAdd{izzet.sakalli@emu.edu.tr}
\emailAdd{ahmadbadawi@ahu.edu.jo}
\emailAdd{abdelmalekbouzenada@gmail.com}

\abstract{
This paper investigates scalar perturbations and quasinormal modes (QNMs) associated with cylindrical black holes constructed within the frameworks of $f(\mathcal{R})$-gravity and Ricci-Inverse ($\mathcal{RI}$) gravity. Moreover, we study the modified Hawking radiation in these black hole solutions and analyze the effects of coupling constants.  These modified theories, which extend general relativity by introducing higher-order curvature corrections and additional geometric terms, provide a rich platform for exploring deviations from standard gravitational physics. The study begins by revisiting the cylindrical black holes in these modified gravity theories, where the effective cosmological constants respectively, are represented by $\Lambda_m^{f(\mathcal{R})}$ and $\Lambda_m^{\mathcal{RI}}$ related to the coupling constants unique to each framework. Afterwards, the QNMs, intrinsic damped oscillations of the black hole space-time, are analyzed to probe the stability of the system, with the effective potential $V$ revealing the impact of the modified gravity parameters. Additionally, the thermodynamic properties of the black holes are examined through the lens of the Generalized Uncertainty Principle (GUP), which introduces quantum corrections to Hawking radiation. The GUP-modified Hawking temperature and entropy are derived, demonstrating significant deviations from classical results and highlighting the quantum gravitational effects in these modified frameworks. By linking QNMs, thermodynamics, and quantum corrections, this work not only deepens the understanding of modified gravity theories but also offers potential observational pathways to test their validity.}


\begin{document}
\maketitle
\flushbottom

\section{Introduction}

For decades, black holes have captivated scientists as some of the most fascinating and enigmatic entities in the universe, primarily due to their unique role in Einstein's theory of gravity. While many black holes are believed to form from the gravitational collapse of massive stars (see Refs. \cite{b1,b2,b3,b4,b5}), others, such as primordial black holes, may emerge from density fluctuations during cosmic inflation or from early-universe topological defects. Observational breakthroughs, including measurements of black hole spins in X-ray binaries, gravitational wave detections from binary mergers via LIGO, the imaging of the supermassive black hole in galaxy M87 by the Event Horizon Telescope \cite{b6,b7,b8,b9}, and the discovery of star-black hole binary systems through radial velocity measurements \cite{b10}, strongly corroborate the existence of black holes. These advances not only deepen our understanding of gravity but also provide invaluable insights into the properties and dynamics of black holes.

Stephen Hawking's groundbreaking contributions have significantly advanced our knowledge of black hole physics. At the core of a black hole lies a singularity \cite{a4}, a region of infinite density where classical physical laws break down \cite{a5}. Surrounding this singularity is the event horizon \cite{a6}, a boundary beyond which not even light can escape the black hole’s immense gravitational pull \cite{a7}. Black holes are fundamentally characterized by three parameters: (i) mass \cite{a9}, (ii) electric charge \cite{a10}, and (iii) angular momentum \cite{a11}. Hawking's pioneering work explored quantum phenomena at the event horizon, leading to the concept of Hawking radiation \cite{a15}. This discovery revealed that black holes are not entirely ``black'' but instead emit radiation due to quantum effects, gradually losing mass over time. Such findings connect thermodynamics with quantum mechanics by introducing the notions of black hole temperature ($T_{H}$) and entropy ($S$) \cite{a16,a17}. Additionally, Hawking's research linked black holes to gravitational waves \cite{a18} and quantum states \cite{a19}, laying the foundation for efforts to unify quantum mechanics and general relativity (GR) \cite{a20,a21,a22,a23}. These theories continue to inspire contemporary investigations into the quantum nature of spacetime \cite{a24,a25}.

Among the modified theories of gravity, $f(\mathcal{R})$-gravity has garnered significant attention \cite{c1,c2,c3}. This framework generalizes the Ricci scalar $R$ in the Einstein-Hilbert action to an arbitrary function $f(\mathcal{R})$ \cite{c4,c5}, introducing corrections that can explain phenomena such as the accelerated expansion of the universe. While $f(\mathcal{R})$-gravity holds promise, it faces theoretical and observational challenges \cite{c6,c7,c8}. Furthermore, the interplay between black hole physics and spacetime modifications is evident in scenarios involving Lorentz violations, where the geometry near the event horizon influences the Hawking radiation spectrum and temperature. Such deviations from GR predictions provide potential avenues for constraining Lorentz-violating effects through black hole thermodynamics \cite{h1,h2,h3,h4,h5}.

QNMs have long been recognized as essential tools for probing the stability of spacetime configurations, such as black holes or branes. QNMs, which manifest as damped oscillations with discrete spectra, are intrinsic to the black hole's structure, independent of the initial perturbations \cite{d7}. The study of QNMs, initiated in \cite{d4,d5,d6}, has illuminated their significance in characterizing black hole mass, angular momentum \cite{d8,d9}, and stability \cite{d8}. Moreover, in the context of AdS/CFT correspondence, QNMs elucidate the thermalization rates in the boundary theory \cite{d10}. Beyond stability analysis, QNMs contribute to our understanding of Hawking radiation \cite{d11} and offer insights into quantizing gravity. The boundary conditions governing QNMs-ingoing waves at the event horizon and theory-specific conditions at infinity-underscore their role in black hole physics and related theoretical frameworks \cite{d1}.

Recently, $\mathcal{RI}$- gravity has emerged as a compelling alternative framework within the realm of modified gravity theories, offering new insights into some of the most perplexing phenomena in theoretical physics, including causality violations, wormholes, and stellar structures \cite{Amendola:2020qho,Malik:2024zhb,arxiv,Ahmed:2024jsh,Moreira:2024mkb,Ahmed:2024jcb,Mustafa:2023mls,Malik:2023yaj,ref4,ref5,ref8,ref9,plb,meer,AM1,AM2,AM3,MFS2,ijtp,aop,EPJP,NA,NPB,EPJC,EPJC3,EPJC4,JCAP}. This class of theories generalizes the standard Einstein-Hilbert action by incorporating higher-order curvature terms and other geometric quantities, leading to modifications of the field equations that accommodate rich and diverse cosmological and astrophysical implications. Among these, the study of cylindrical black holes \cite{arxiv} plays a crucial role, as they provide simplified yet insightful models to explore the effects of these modifications in a highly symmetric spacetime. Unlike the more commonly studied spherical black holes, cylindrical configurations inherently possess an anisotropic structure, which enables a detailed investigation of directional dependencies and dynamical behaviors unique to modified gravity theories. QNMs associated with cylindrical black holes further enhance their utility as probes of stability, as these damped oscillations are highly sensitive to the underlying spacetime geometry and the parameters of the $\mathcal{RI}$-gravity model. Moreover, cylindrical black holes offer an ideal framework for studying thermodynamic properties, such as Hawking radiation and entropy, particularly when quantum corrections like those introduced by the GUP are considered. These investigations not only shed light on the deviations from GR but also highlight the intricate interplay between modified gravity, black hole thermodynamics, and quantum phenomena. By examining these systems, $\mathcal{RI}$-gravity provides a robust platform for addressing unresolved questions about the nature of spacetime, the dynamics of exotic matter, and the potential observational signatures that could bridge the gap between theoretical predictions and empirical evidence. Consequently, the study of cylindrical black holes in this context serves as a stepping stone toward a deeper understanding of gravitational physics beyond Einstein's framework.

The paper is organized as follows: In Sec.~\ref{sec2}, we present the cylindrical black hole solution within $f(\mathcal{R})$ and $\mathcal{RI}$-gravity frameworks. Sec.~\ref{sec3} focuses on the QNMs, analyzing the effective potential and their stability. In Sec.~\ref{sec4}, we explore the impact of the GUP on the Hawking radiation and thermodynamic properties of the black holes. Finally, Sec.~\ref{sec5} concludes the paper, summarizing the results and discussing potential future directions.

\section{BH solutions within $f(\mathcal{R})$ and $\mathcal{RI}$-gravity theories} \label{sec2}

In this section, we consider a cylindrical black hole solution formulated within GR, in the context of modified gravity theory. Therefore, we begin this section by introducing this BH solution in cylindrical coordinates ($t, r, \varphi, z$) given by \cite{JPSL}
\begin{equation}
ds^2=-f(r)\,dt^2+\frac{dr^2}{f(r)}+r^2\,d\varphi^2+\alpha^2\,r^2\,dz^2,\quad\quad f(r)=\left(\alpha^2\,r^2-\frac{4\,M}{\alpha\,r} \right),\label{a1}
\end{equation}
where the metric tensor $g_{\mu\nu}$ and its contravairant form are given by $(x^0=t, x^1=r, x^2=\varphi, x^3=z)$
\begin{eqnarray}
g_{\mu\nu}=\left(\begin{array}{cccc}
     -f(r) & 0 & 0 & 0 \\
     0 & \frac{1}{f(r)} & 0 & 0\\
     0 & 0 & r^2 & 0\\
     0 & 0 & 0 & \alpha^2\,r^2 
\end{array}\right),\quad\quad 
g^{\mu\nu}=\left(\begin{array}{cccc}
     -\frac{1}{f(r)} & 0 & 0 & 0\\
     0 & f(r) & 0 & 0\\
     0 & 0 & \frac{1}{r^2} & 0\\
     0 & 0 & 0 & \frac{1}{\alpha^2\,r^2}
\end{array}\right).\label{a2}
\end{eqnarray}
Here $M$ denotes mass of BH, $\alpha>0$ is a positive constant, and the coordinates are in the ranges $- \infty <t < \infty$, $r \geq 0$, $\varphi \in[0, 2\,\pi)$, and $- \infty < z < \infty$ called the temporal, radial, angular and the axial coordinates, respectively.

This space-time (\ref{a1}) satisfied the Einstein vacuum field equations with a negative cosmological constant $(\Lambda<0)$ given by
\begin{equation}
    R_{\mu\nu}=\Lambda\,g_{\mu\nu},\quad\quad \mathcal{R}=4\,\Lambda,\quad\quad \Lambda=-3\,\alpha^2.\label{a3}
\end{equation}

Below, we review this black hole (BH) solution in the context of modified gravity theories. Specifically, we focus on two prominent theories: $f(\mathcal{R})$-gravity and $\mathcal{RI}$-gravity. These modified theories extend GR by altering the gravitational Lagrangian, introducing additional degrees of freedom, and potentially offering insights into the nature of dark energy, dark matter, and the early universe.

Our main motivation is to investigate the scalar perturbations around the modified black hole solutions obtained in the gravity theories mentioned earlier. Scalar fields play a crucial role in modified gravity theories and can lead to deviations from the standard GR predictions. Finally, we study the Hawking temperature, which is a key feature of black hole thermodynamics. The Hawking temperature, associated with the emission of black hole radiation, is influenced by the spacetime geometry and the gravitational modifications consider in the current study. By calculating the Hawking temperature for the black holes in $f(\mathcal{R})$-gravity and $\mathcal{RI}$-gravity, we can compare the results with those from GR and gain insights into how these modifications affect the thermodynamic properties of black holes.

\vspace{0.1cm}
\begin{center}
    {\bf Case A: $f(\mathcal{R})$-gravity theory}
\end{center}
\vspace{0.1cm}

In the framework of $f(\mathcal{R})$-gravity, the action is modified by an arbitrary function of the Ricci scalar $\mathcal{R}$, which leads to modified field equations. This introduces a richer structure to the gravitational dynamics, as the function $f(\mathcal{R})$ allows for deviations from the standard Einstein-Hilbert action.

The Einstein-Hilbert action in $f(\mathcal{R})$-gravity theory is given by
\begin{equation}\label{CC1}
    S = \int \mathrm{d}x^4 \sqrt{-g}\,\left(f(\mathcal{R})-2\,\Lambda_m \right)+ S_M,\quad\quad\quad \mathcal{R}=g_{\mu\nu}\,R^{\mu\nu}.
\end{equation}

By varying the action (\ref{CC1}) with respect to the metric tensor $g_{\mu\nu}$, the modified field equations in $f(\mathcal{R})$-gravity are given by  
\begin{equation}\label{CC2}
    -\frac{1}{2}\,f(\mathcal{R})\,g^{\mu \nu} + f_{\mathcal{R}} \, R^{\mu \nu}+g^{\mu \nu} \nabla ^2 \, f_{\mathcal{R}}-\nabla ^{\mu} \nabla^{\nu}\, f_{\mathcal{R}}+ \Lambda_m\, g^{\mu \nu} =\mathcal{T}^{\mu \nu},
\end{equation}
where $\Lambda_m$ is the effective cosmological constant in this modified gravity.

Let us consider the function $f(\mathcal{R})$ to be the following form:
\begin{equation} \label{CC3}
    f(\mathcal{R}) =\mathcal{R}+\alpha_1\,\mathcal{R}^2+\alpha_2\,\mathcal{R}^3+\alpha_3\,\mathcal{R}^4+\alpha_4\,\mathcal{R}^5, 
\end{equation}
where $\alpha_i\quad (i=1,..,4)$ are the coupling constants. 

By solving the modified field equations (\ref{CC2}) using the metric tensor (\ref{a2}), the relations (\ref{a3}), and the function (\ref{CC3}), and performing the necessary simplifications for zero energy-momentum tensor, $\mathcal{T}^{\mu\nu}$, we obtain the effective cosmological constant as follows \cite{arxiv}:
\begin{equation}
    \Lambda^{f(\mathcal{R})}_m=-3\,\alpha^2+432\,\alpha^6\,\alpha_2-10368\,\alpha^8\,\alpha_3+186624\,\alpha^{10}\,\alpha_4.\label{CC4}
\end{equation}

In terms of usual cosmological constant $\Lambda$, we can write the effective cosmological constant
\begin{equation}
    \Lambda^{f(\mathcal{R})}_m=\Lambda-16\,\Lambda^3\,\alpha_2-128\,\Lambda^4\,\alpha_3-768\,\alpha_4\,\Lambda^5.\label{CC5}
\end{equation}

The effective cosmological constant $\Lambda_m$ is negative provided the following condition must obey:
\begin{equation}\label{CC6}
    16\,\Lambda^3\,\alpha_2+768\,\alpha_4\,\Lambda^5 <\Lambda-128\,\Lambda^4\,\alpha_3.
\end{equation}

Now, we consider a general function of $f(\mathcal{R})$ given by the following form:
\begin{equation}\label{CC7}
    f(\mathcal{R})=\mathcal{R}+\alpha_k\,\mathcal{R}^{k+1},\quad k=1,2,....n.
\end{equation}

Solving the modified field equations (\ref{CC2}) for zero energy-momentum tensor, $\mathcal{T}^{\mu\nu}=0$, and after simplification, we obtain \cite{arxiv}
\begin{equation}
    \Lambda^{f(\mathcal{R})}_m=-3\,\alpha^2-3\,\alpha^2\,(1-k)\,\alpha_k\,(-12\,\alpha^2)^{k}.  \label{CC8}
\end{equation}
In terms of the usual cosmological constant $\Lambda$, we can rewrite the effective cosmological constant as follows:
\begin{equation}\label{CC9}
    \Lambda^{f(\mathcal{R})}_m=\Lambda+\alpha_k\,(1-k)\,4^k\,\Lambda^{k+1},\quad (k=1,2,....n).
\end{equation}

The BH solution (\ref{a1}) thus is a valid solution in $f(\mathcal{R})$-gravity theory. In terms of the effective cosmological constant $\Lambda^{f(\mathcal{R})}_m$, the BH solution in $f(\mathcal{R})$ gravity framework is described by the following line-element
\begin{eqnarray}
ds^2=\left[\frac{\Lambda^{f(\mathcal{R})}_m}{3}r^2+\frac{4\,M}{\sqrt{-\frac{\Lambda^{f(\mathcal{R})}_m}{3}}\,r}\right]dt^2-\frac{dr^2}{\left[\frac{\Lambda^{f(\mathcal{R})}_m}{3}\,r^2+\frac{4\,M}{\sqrt{-\frac{\Lambda^{f(\mathcal{R})}_m}{3}}\,r}\right]}+r^2\,d\varphi^2-\frac{\Lambda^{f(\mathcal{R})}_m}{3}\,r^2\,dz^2,\label{CC10}
\end{eqnarray}
where $\Lambda^{f(\mathcal{R})}_m<0$ is given in Eq. (\ref{CC5}) and (\ref{CC9}).

\vspace{0.1cm}
\begin{center}
    {\bf Case B: $\mathcal{RI}$-gravity theory}
\end{center}
\vspace{0.1cm}

Now, we study the BH solution (\ref{a1}) in the framework of $\mathcal{RI}$-gravity theory. In this new gravitational theory, the Einstein-Hilbert action of GR is modified by a function of the Ricci scalar $(\mathcal{R})$, the anti-curvature scalar $\mathcal{A}=g_{\mu\nu}\,A^{\mu\nu} \neq \mathcal{R}^{-1}$, and the anti-curvature tensor $A^{\mu\nu}=\mathcal{R}^{-1}_{\mu\nu}$.   

Therefore, the action in the framework of $\mathcal{RI}$-gravity is described by 
\begin{equation}\label{B1}
    \mathcal{S}=\int \mathrm{d}x^4 \sqrt{-g}\,\left(f(\mathcal{R},\mathcal{A},A^{\mu \nu}A_{\mu \nu})-2\,\Lambda_m\right)+ S_m.
\end{equation}

By varying the action (\ref{B1}) with respect to the metric tensor $g_{\mu\nu}$, the modified field equations in Class-{\bf III} model of $\mathcal{RI}$-gravity are given by  
\begin{equation}\label{B2}
    -\frac{1}{2}\,f\,g^{\mu\nu}+f_{\mathcal{R}}\,R^{\mu \nu}-f_{\mathcal{A}}\,A^{\mu\nu}-2\,f_{A^2}\,A^{\rho \nu }\,A^{\mu}_{\rho}+P^{\mu\nu}+M^{\mu\nu}+U^{\mu\nu}+ \Lambda_m\,g^{\mu\nu}=\mathcal{T}^{\mu\nu},
\end{equation}
where 
\begin{eqnarray}
    P^{\mu\nu}&=&g^{\mu\nu}\,\nabla ^2\,f_{\mathcal{R}}-\nabla^{\mu} \nabla^{\nu}\,f_{\mathcal{R}},\label{B3}\\ 
    M^{\mu\nu}&=&g^{\rho\mu}\,\nabla_{\alpha}\,\nabla_{\rho }(f_{\mathcal{A}}\, A_{\sigma}^{\alpha}\,A^{\nu\sigma})-\frac{1}{2}\,\nabla^2 (f_{\mathcal{A}}\,A^{\mu}_{\sigma}\,A^{\nu\sigma})-\frac{1}{2}\,g^{\mu\nu}\, \nabla_{\alpha}\,\nabla_{ \rho}(f_{\mathcal{A}}\,A_{\sigma}^{\alpha}\,A^{\rho \sigma}),\label{B4}\\
 \nonumber  U^{\mu\nu}&=&g^{\rho\nu}\,\nabla_{\alpha}\,\nabla_{\rho}(f_{A^2}\,A_{\sigma\kappa}\,A^{\sigma\alpha}\,A^{\mu\kappa})-\nabla^{2}(f_{A^2}\,A_{\sigma \kappa}\,A^{\sigma\mu}\,A^{\nu\kappa})
  -g^{\mu\nu}\,\nabla_{\alpha}\,\nabla_{\rho}(f_{A^2}\,A_{\sigma \kappa}\,A^{\sigma\alpha}\,A^{\rho\kappa})\nonumber\\
  &+&2\,g^{\rho\nu}\,\nabla_{\rho}\,\nabla_{\alpha}(f_{A^2}\,A_{\sigma \kappa}\,A^{\sigma\mu}\,A^{\alpha\kappa})
   - g^{\rho\nu}\,\nabla_{\alpha}\,\nabla_{\rho}(f_{A^2}\,A_{\sigma \kappa}\,A^{\sigma\mu}\,A^{\alpha\kappa}).\label{B5} 
\end{eqnarray}
Here $f_{\mathcal{R}},f_{\mathcal{A}}$ are defined earlier and $f_{A^2}=\partial f/\partial (A^{\mu\nu}\,A_{\mu\nu})$.

Let's consider the function $f$ in this Class-{\bf III} model to be the following form:
\begin{equation} \label{B6}
    f(\mathcal{R},\mathcal{A}, A^{\mu \nu}\,A_{\mu \nu}) =\mathcal{R}+\alpha_1\,\mathcal{R}^2+\alpha_2\,\mathcal{R}^3+\beta_1 \, \mathcal{A} +\beta_2 \, \mathcal{A}^2+\gamma\,A^{\mu \nu}\,A_{\mu \nu} \, 
\end{equation}
with $\alpha_i,\beta_i$ and $\gamma$ being arbitrary constants. 

After solving the modified field equations (\ref{B2}) for zero energy-momentum tensor, $\mathcal{T}^{\mu\nu}=0$, we find \cite{arxiv}
\begin{equation}
\Lambda^{\mathcal{RI}}_m=-3\,\alpha^2+432\,\alpha^6\,\alpha_2-\frac{\beta_1}{\alpha^2}+\frac{16\,\beta_2}{9\,\alpha^4}+\frac{4\,\gamma}{9\,\alpha^4}.\label{B7}
\end{equation}

In terms of usual cosmological constant $\Lambda$, we can write the effective cosmological constant as follows:
\begin{equation}
    \Lambda^{\mathcal{RI}}_m=\Lambda-16\,\Lambda^3\,\alpha_2+\frac{3\,\beta_1}{\Lambda}+\frac{16\,\beta_2}{\Lambda^2}+\frac{4\,\gamma}{\Lambda^2}.\label{B8}
\end{equation}

The BH solution (\ref{a1}) in terms of usual cosmological constant ($\Lambda$) within the framework of GR is described by the following line-element
\begin{eqnarray}
ds^2=-\left(-\frac{\Lambda}{3}\,r^2-\frac{4\,M}{\sqrt{-\frac{\Lambda}{3}}\,r}\right)\,dt^2+\frac{dr^2}{\left(-\frac{\Lambda}{3}\,r^2-\frac{4\,M}{\sqrt{-\frac{\Lambda}{3}}\,r}\right)}+r^2\,d\varphi^2-\frac{\Lambda}{3}\,r^2\,dz^2.\label{B10}
\end{eqnarray}

The BH solution (\ref{a1}) in terms of effective cosmological constant $\Lambda_m$ within the framework of $\mathcal{RI}$-gravity is therefore described by the following line-element
\begin{eqnarray}
ds^2=\left(\frac{\Lambda^{\mathcal{RI}}_m}{3}\,r^2+\frac{4\,M}{\sqrt{-\frac{\Lambda^{\mathcal{RI}}_m}{3}}\,r}\right)\,dt^2-\frac{dr^2}{\left(\frac{\Lambda^{\mathcal{RI}}_m}{3}\,r^2+\frac{4\,M}{\sqrt{-\frac{\Lambda^{\mathcal{RI}}_m}{3}}\,r}\right)}+r^2\,d\varphi^2-\frac{\Lambda^{\mathcal{RI}}_m}{3}\,r^2\,dz^2,\label{B11}
\end{eqnarray}
where $\Lambda^{\mathcal{RI}}_m<0$ is given in Eq. (\ref{B8}).

In the subsequent sections, we first examine the scalar perturbations of the modified black hole solutions, and then investigate the Hawking temperature. We analyze how the modifications introduced by the gravity theories affect these quantities, and compare the results with those predicted by GR, highlighting the shifts and deviations caused by the modified gravitational frameworks.

\section{Scalar Perturbations and the Effective Potential: QNMs} \label{sec3}

In this section, we investigate the perturbations of a massless spin-0 scalar field by solving the Klein-Gordon equation within the context of the modified gravity theories introduced in the previous section. The scalar field, denoted $\Phi$, is considered to be massless, and its dynamics is governed by the general covariant Klein-Gordon equation. This equation serves as the fundamental equation of motion for spin-0 massless scalar field in a curved spacetime and takes the form given by \cite{LL}:
\begin{equation}
    \frac{1}{\sqrt{-g}}\,\partial_{\mu}\,\left(\sqrt{-g}\,g^{\mu\nu}\,\partial_{\nu}\Phi\right)=0,\label{C1}
\end{equation}
where $g_{\mu\nu}$ is the metric tensor of the spacetime, and $g$ is its determinant. In the context of modified gravity theories, such as $f(\mathcal{R})$-gravity or $\mathcal{RI}$-gravity, the background spacetime and the equation of motion can be altered due to the modifications to the gravitational action. These modifications can affect the propagating behavior of the scalar field, and we aim to study how the field behaves under perturbations in these modified frameworks.

Given that the background space-time (\ref{CC10}) and (\ref{B11}) is cylindrical symmetric and static, we can express the massless scalar field function $\Phi(t,r,\varphi,z)$ as a decomposition in terms of $\mathrm{R}(r)$ as follows:
\begin{equation}
    \Phi(t,r,\varphi,z)=\exp(-i\,\omega\,t)\,\exp(i\,m\,\varphi)\,\exp(i\,k\,z)\,\frac{\mathrm{R}(r)}{r},\label{C2}
\end{equation}
where $\omega$ is the QNMs frequency, $m$ takes the natural number and $k$ is an arbitrary constant.

The modified BH solutions obtained earlier in the framework of $f(\mathcal{R})$ and $\mathcal{RI}$-gravity together can be expressed as
\begin{eqnarray}
ds^2=\left(\frac{\Lambda_m}{3}\,r^2+\frac{4\,M}{\sqrt{-\frac{\Lambda_m}{3}}\,r}\right)\,dt^2-\frac{dr^2}{\left(\frac{\Lambda_m}{3}\,r^2+\frac{4\,M}{\sqrt{-\frac{\Lambda_m}{3}}\,r}\right)}+r^2\,d\varphi^2-\frac{\Lambda_m}{3}\,r^2\,dz^2,\label{special}
\end{eqnarray}
Now, we calculate the determinant of the metric tensor (\ref{special}) and it is given by
\begin{equation}
    \sqrt{-g}=\sqrt{-\frac{\Lambda_m}{3}}\,r^2,\label{C3}
\end{equation}
where $\Lambda_m \to \Lambda^{f(\mathcal{R})}_m$ as given in Eq. (\ref{CC5}) and (\ref{CC9}), and $\Lambda_m \to \Lambda^{\mathcal{RI}}_m$ as specified in Eq. (\ref{B8}).  

With these, equation (\ref{C1}) can be expressed as
\begin{equation}
f^2(r)\,\partial^2_{r}\,\mathrm{R}+f(r)\,f'(r)\,\partial_{r}\,\mathrm{R}+\left(\omega^2-\frac{\iota^2}{r^2}\,f(r)-\frac{f\,f'}{r}\right)\,\mathrm{R}(r)=0,\label{C4}
\end{equation}
where prime denotes ordinary derivative w. r. t. $r$ and
\begin{equation}
    \iota=\sqrt{m^2-\frac{3\,k^2}{\Lambda_m}},\quad\quad\quad f(r)=\left(-\frac{\Lambda_m}{3}\,r^2-\frac{4\,M}{\sqrt{-\frac{\Lambda_m}{3}}\,r} \right).\label{C5}
\end{equation}

We now introduce a new coordinate called the tortoise coordinate, $r_*$, for the modified BH solutions (\ref{special}) defined by
\begin{equation}
    dr_{*}=\frac{dr}{f(r)},\quad\quad \partial_{r_*}=f(r)\,\partial_{r}\label{C6}
\end{equation}
One can rewrite the above differential equation (\ref{C4}) as follows:
\begin{equation}
    \partial^2_{r_*}\,\mathrm{R}+\left(\omega^2-V\right)\,\mathrm{R}=0,\label{C7}
\end{equation}
where the effective potential of the scalar perturbations is given by
\begin{eqnarray}
    V(r)&=&\left(\frac{\iota^2}{r^2}+\frac{f'(r)}{r}\right)\,f(r)\nonumber\\
    &=&\left[\frac{1}{r^2}\,\left(m^2-\frac{3\,k^2}{\Lambda_m}\right)-\frac{2\,\Lambda_m}{3}+\frac{4\,M}{\sqrt{-\frac{\Lambda_m}{3}}\,r^3} \right]\,\left(-\frac{\Lambda_m}{3}\,r^2-\frac{4\,M}{\sqrt{-\frac{\Lambda_m}{3}}\,r} \right).\label{C8} 
\end{eqnarray}

In the context of $f(\mathcal{R})$-gravity, the effective potential will be 
\small
\begin{eqnarray}
    V_{f(\mathcal{R})}(r)&=&\Bigg[\frac{1}{r^2}\left(m^2-\frac{3\,k^2}{(\Lambda-16\,\Lambda^3\,\alpha_2-128\,\Lambda^4\,\alpha_3-768\,\alpha_4\,\Lambda^5)}\right)\nonumber\\
    &-&\frac{2\,(\Lambda-16\,\Lambda^3\,\alpha_2-128\,\Lambda^4\,\alpha_3-768\,\alpha_4\,\Lambda^5)}{3}
    +\frac{4\,M}{r^3\sqrt{-\frac{(\Lambda-16\,\Lambda^3\,\alpha_2-128\,\Lambda^4\,\alpha_3-768\,\alpha_4\,\Lambda^5)}{3}}} \Bigg]\times\nonumber\\
    &&\left[-\frac{(\Lambda-16\,\Lambda^3\,\alpha_2-128\,\Lambda^4\,\alpha_3-768\,\alpha_4\,\Lambda^5)}{3}r^2-\frac{4\,M}{r\sqrt{-\frac{(\Lambda-16\,\Lambda^3\,\alpha_2-128\,\Lambda^4\,\alpha_3-768\,\alpha_4\,\Lambda^5)}{3}}} \right].\label{C9} 
\end{eqnarray}
\normalsize

And in the framework of $\mathcal{RI}$-gravity, it will be
\begin{eqnarray}
    V_{\mathcal{RI}}(r)&=&\Bigg[\frac{1}{r^2}\left(m^2-\frac{3\,k^2}{\left(\Lambda-16\,\Lambda^3\,\alpha_2+\frac{3\,\beta_1}{\Lambda}+\frac{16\,\beta_2}{\Lambda^2}+\frac{4\,\gamma}{\Lambda^2}\right)}\right)\nonumber\\
    &-&\frac{2\,\left(\Lambda-16\,\Lambda^3\,\alpha_2+\frac{3\,\beta_1}{\Lambda}+\frac{16\,\beta_2}{\Lambda^2}+\frac{4\,\gamma}{\Lambda^2}\right)}{3}
    +\frac{4\,M}{r^3\sqrt{-\frac{\left(\Lambda-16\,\Lambda^3\,\alpha_2+\frac{3\,\beta_1}{\Lambda}+\frac{16\,\beta_2}{\Lambda^2}+\frac{4\,\gamma}{\Lambda^2}\right)}{3}}} \Bigg]\times\nonumber\\
    &&\left[-\frac{\left(\Lambda-16\,\Lambda^3\,\alpha_2+\frac{3\,\beta_1}{\Lambda}+\frac{16\,\beta_2}{\Lambda^2}+\frac{4\,\gamma}{\Lambda^2}\right)}{3}r^2-\frac{4\,M}{r\sqrt{-\frac{\left(\Lambda-16\,\Lambda^3\,\alpha_2+\frac{3\,\beta_1}{\Lambda}+\frac{16\,\beta_2}{\Lambda^2}+\frac{4\,\gamma}{\Lambda^2}\right)}{3}}} \right].\label{C10} 
\end{eqnarray}

\begin{figure}[ht!]
    \centering
    \includegraphics[width=0.45\linewidth]{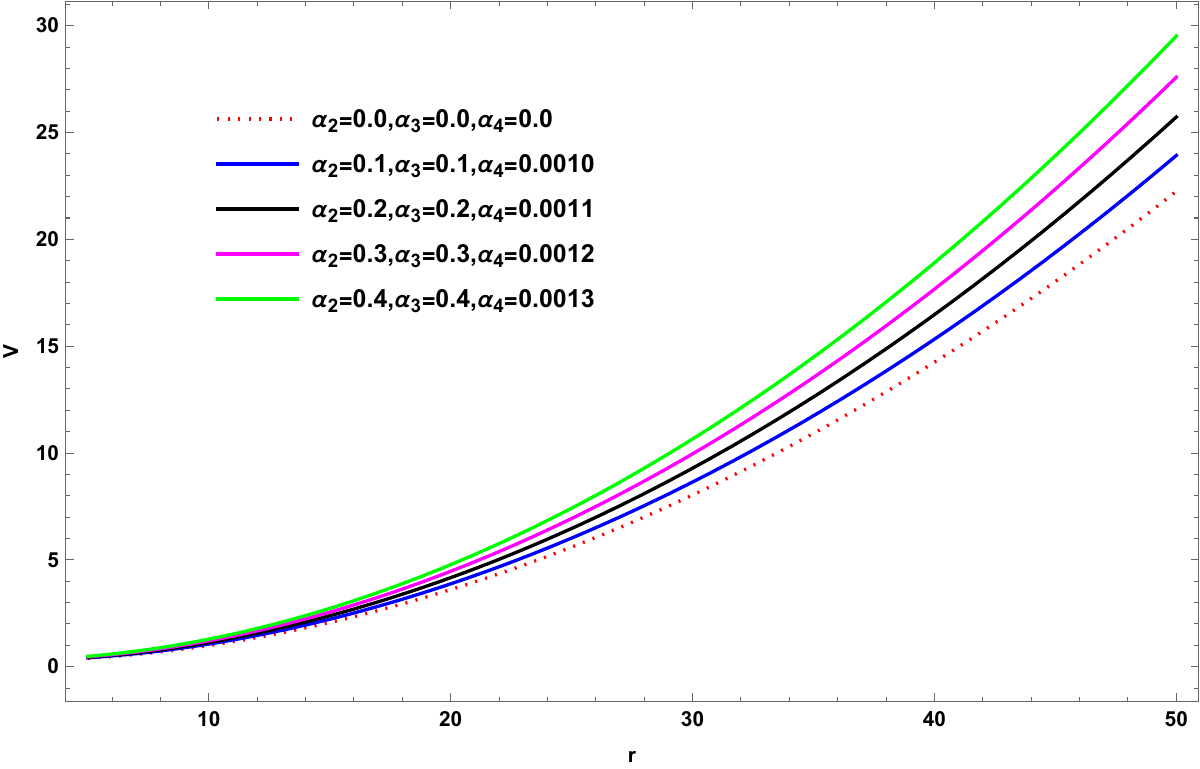}\quad\quad
    \includegraphics[width=0.45\linewidth]{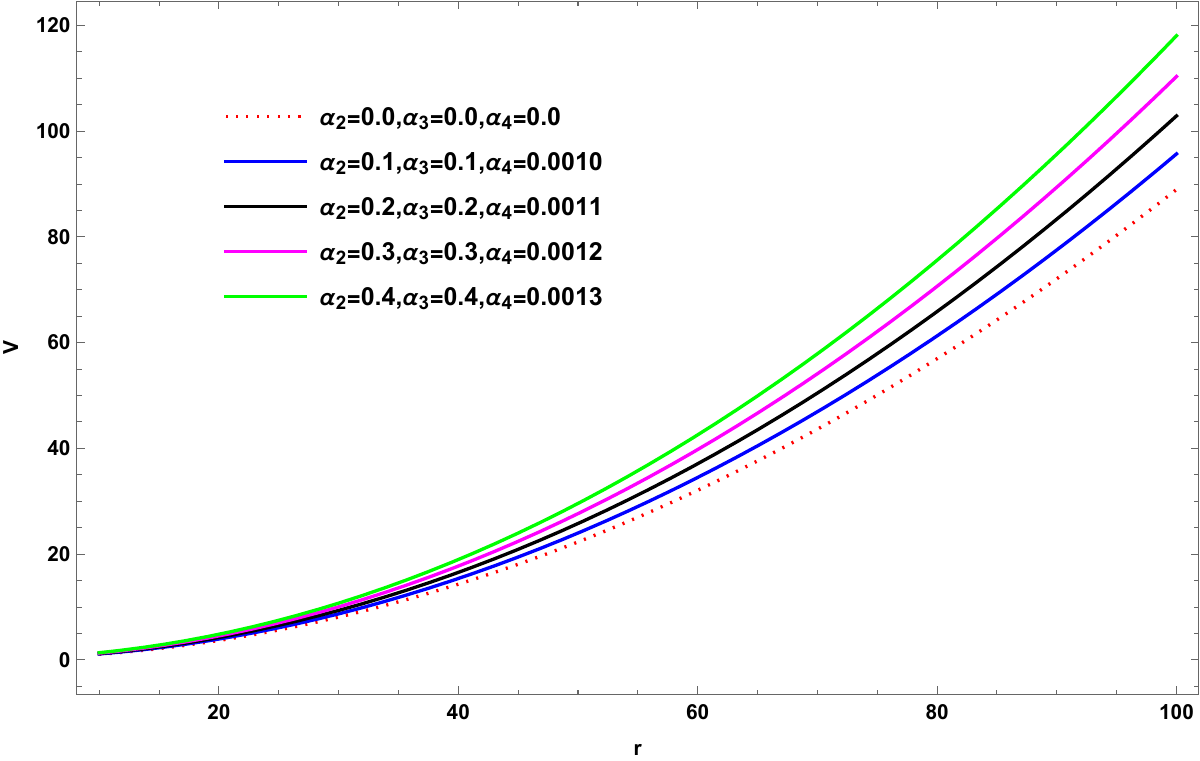}
    \caption{A comparison of the effective potential (\ref{C9}) with GR case for scalar perturbations is plotted, considering cases: $m=0$ (left panel) and $m=1$ (right panel). Here $M=1$, $\Lambda=-0.2$, and $k=0.01$.}
    \label{fig:1}
    \hfill\\
    \includegraphics[width=0.45\linewidth]{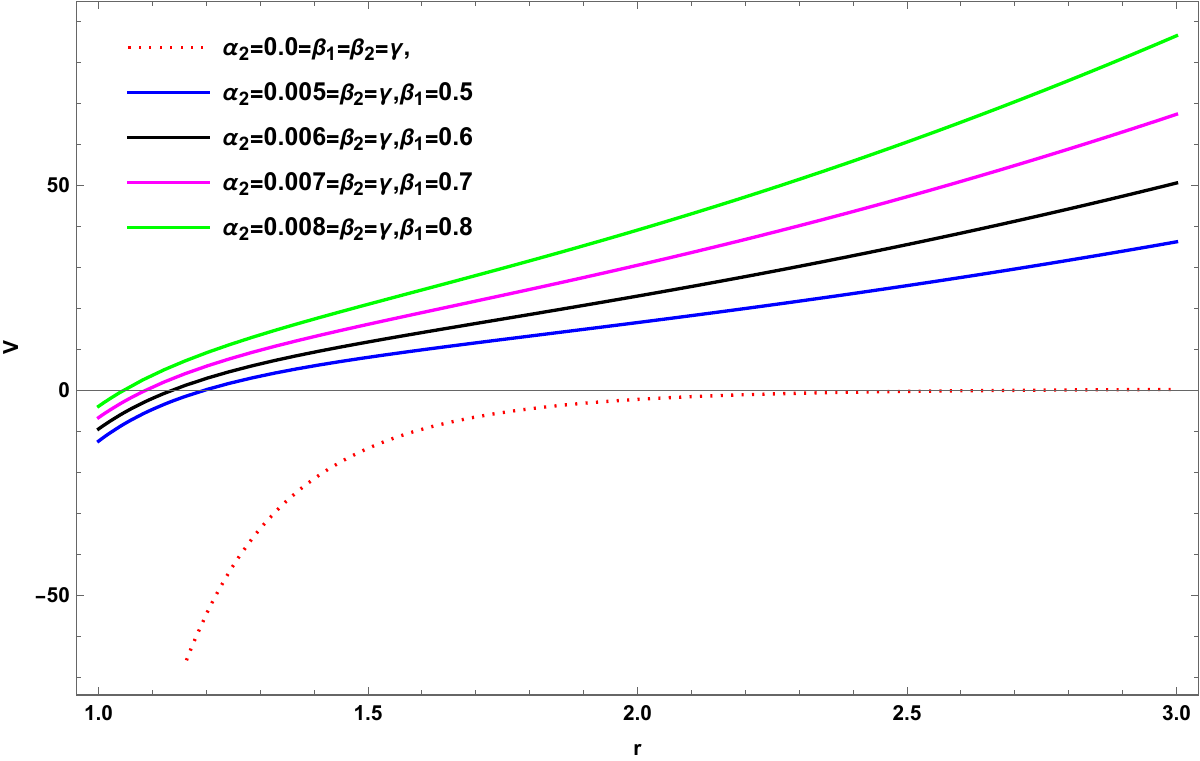}\quad\quad
    \includegraphics[width=0.45\linewidth]{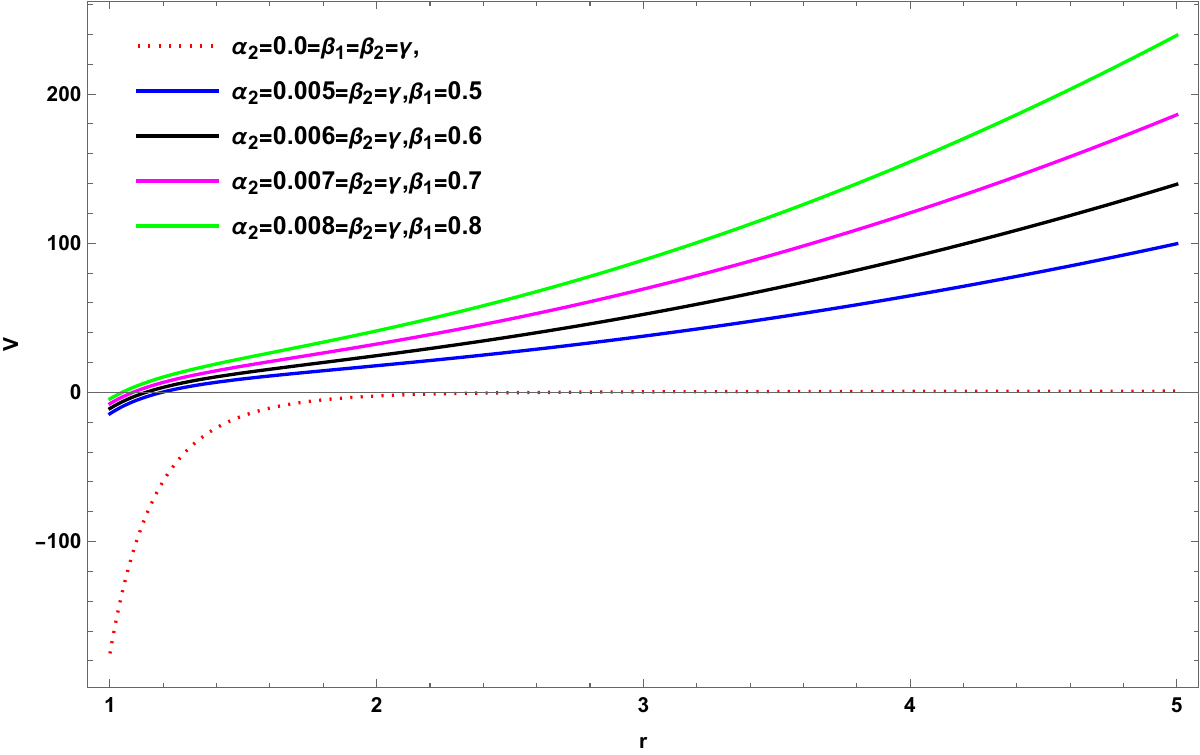}
    \caption{A comparison of the effective potential (\ref{C10}) with the GR case for scalar perturbations is plotted, considering $m=0$ (left panel) and $m=1$ (right panel). Here $M=1$, $\Lambda=-0.2$, and $k=0.1$.}
    \label{fig:2}
\end{figure}

From the expressions in Eqs. (\ref{C9}) and (\ref{C10}), it is evident that the effective potential governing the scalar perturbations is significantly influenced by the coupling constants present in the modified gravitational theories. In the context of $f(\mathcal{R})$-gravity, the effective potential is modulated by the constants $\alpha_2$, $\alpha_3$, and $\alpha_4$, which arise from the functional form of the gravitational action and the associated field equations. These constants determine the strength and nature of the interaction between the scalar field and the modified curvature terms, thereby affecting the propagation of scalar fields.

Similarly, in the framework of $\mathcal{RI}$-gravity, the effective potential is influenced by a different set of parameters: $\alpha_2$, $\beta_1$, $\beta_2$, and $\gamma$. These constants arise from the specific modification used in $\mathcal{RI}$-gravity. The interplay between these parameters governs how the scalar field interacts with the modified gravitational background, which affects the dynamics of scalar perturbations in this model.

We have generated Figure \ref{fig:1}, which illustrates the effective potential for scalar perturbations within the framework of $f(\mathcal{R})$ gravity, considering various values of the coupling constants. The figure shows how the effective potential evolves with changes in the values of the coupling constants $\alpha_2$, $\alpha_3$, and $\alpha_4$. As these constants increase, the potential gradually rises, indicating that the strength of the scalar perturbations is affected by the modifications to the gravitational action.

Furthermore, we compare this potential to the standard GR case by setting $\alpha_2 = 0$, $\alpha_3 = 0$, and $\alpha_4 = 0$, effectively reducing $f(\mathcal{R})$ gravity to the GR limit. This comparison highlights the differences between the modified gravity theory and GR. In the figure, the red dotted lines represent the effective potential in the GR case, while the solid colored lines correspond to the potential in the $f(\mathcal{R})$ gravity framework with non-zero coupling constants. The distinction between these two cases provides insight into how the modifications to gravity, introduced through the coupling constants, alter the behavior of scalar perturbations in a cosmological context.

Similarly, in Figure \ref{fig:2}, we present the effective potential for scalar perturbations within the framework of $\mathcal{RI}$-gravity, considering various values of the coupling constants. The figure illustrates how the effective potential changes with variations in the coupling constants $\alpha_2$, $\beta_1$, $\beta_2$, and $\gamma$. As these constants increase, the potential gradually rises, suggesting that the strength and nature of the scalar perturbations are influenced by the modifications to the gravitational action introduced by these constants.

Additionally, we compare this potential with the standard GR case by setting $\alpha_2 = 0$, $\beta_1 = 0$, $\beta_2 = 0$, and $\gamma = 0$, which effectively reduces $\mathcal{RI}$ gravity to the GR limit. This comparison emphasizes the differences between the modified $\mathcal{RI}$ gravity theory and GR. In the figure, the red dotted lines represent the effective potential in the GR case, while the solid colored lines correspond to the potential in the $\mathcal{RI}$-gravity framework with non-zero coupling constants.

The distinction between these cases provides valuable insight into how the modifications to gravity, affect the dynamics of scalar perturbations in a cosmological setting. These modifications could lead to observable effects that distinguish $\mathcal{RI}$-gravity and $f(\mathcal{R})$ gravity from GR, particularly in phenomena involving scalar field fluctuations, cosmological perturbations, or the behavior of gravitational waves in modified gravity theories. 

\section{Vectorial Perturbations}

Next, we consider the electromagnetic perturbation, or spin-one vector fields,  requiring the use of the conventional tetrad formalism \cite{b1,AA1},  where a basis $e^{a}_{\mu} ({\bf x})$ is established in relation to the black hole metric tensor $g_{\mu\nu}$. This chosen basis adheres to the conditions
\begin{equation}
    e^{a}_{\mu} ({\bf x})\,e^{\mu}_{b} ({\bf x})=\delta^{a}_{b},\quad\quad e^{a}_{\mu} ({\bf x})\,e^{\nu}_{a} ({\bf x})=\delta^{\nu}_{\mu},\quad\quad g_{\mu\nu}=\eta_{ab}\,e^{a}_{\mu} ({\bf x})\,e^{b}_{\nu} ({\bf x}).\label{D1} 
\end{equation}

Spin-one fields is described by the following Maxwell equations
\begin{equation}
    \frac{1}{\sqrt{-g}}\,\left(g^{\sigma\mu}\,g^{\tau \nu}\,\sqrt{-g}\,\mathcal{F}_{\sigma\tau}\right)_{,\,\nu}=0.\label{D2}
\end{equation}

The spin-one Maxwell field after a straightforward calculation yields
\begin{equation}
    \partial^2_{r_*}\,\psi_e+(\omega^2-\mathcal{V}_e)\,\psi_e=0,\label{D4}
\end{equation}
where the potential has the following explicit form in general relativity:
\begin{equation}
    \mathcal{V}_e=f(r)\,\frac{\iota^2}{r^2}=\frac{1}{r^2}\,\left(-\frac{\Lambda}{3}\,r^2-\frac{4\,M}{\sqrt{-\frac{\Lambda}{3}}\,r} \right)\,\left(m^2-\frac{3\,k^2}{\Lambda}\right).\label{D5}
\end{equation}

In modified gravity theory, we have
\begin{equation}
    \mathcal{V}_e=\frac{1}{r^2}\,\left(-\frac{\Lambda_m}{3}\,r^2-\frac{4\,M}{\sqrt{-\frac{\Lambda_m}{3}}\,r} \right)\,\left(m^2-\frac{3\,k^2}{\Lambda_m}\right).\label{mod}
\end{equation}

Therefore, in the context of $f(\mathcal{R})$-gravity, the effective potential will be 
\begin{eqnarray}
    \mathcal{V}^{f(\mathcal{R})}_e&=&\frac{1}{r^2}\,\left(\frac{(-\Lambda+16\,\Lambda^3\,\alpha_2+128\,\Lambda^4\,\alpha_3+768\,\alpha_4\,\Lambda^5)}{3}\,r^2-\frac{4\,M}{\sqrt{\frac{(-\Lambda+16\,\Lambda^3\,\alpha_2+128\,\Lambda^4\,\alpha_3+768\,\alpha_4\,\Lambda^5)}{3}}\,r} \right)\times\nonumber\\
    &&\left(m^2+\frac{3\,k^2}{(-\Lambda+16\,\Lambda^3\,\alpha_2+128\,\Lambda^4\,\alpha_3+768\,\alpha_4\,\Lambda^5)}\right).\label{D6}
\end{eqnarray}
And in the framework of $\mathcal{RI}$-gravity, it will be
\begin{eqnarray}
    \mathcal{V}^{\mathcal{RI}}_e&=&\frac{1}{r^2}\,\left(\frac{\left(-\Lambda+16\,\Lambda^3\,\alpha_2-\frac{3\,\beta_1}{\Lambda}-\frac{16\,\beta_2}{\Lambda^2}-\frac{4\,\gamma}{\Lambda^2}\right)}{3}\,r^2-\frac{4\,M}{\sqrt{\frac{\left(-\Lambda+16\,\Lambda^3\,\alpha_2-\frac{3\,\beta_1}{\Lambda}-\frac{16\,\beta_2}{\Lambda^2}-\frac{4\,\gamma}{\Lambda^2}\right)}{3}}\,r} \right)\times\nonumber\\
    &&\left(m^2+\frac{3\,k^2}{\left(-\Lambda+16\,\Lambda^3\,\alpha_2-\frac{3\,\beta_1}{\Lambda}-\frac{16\,\beta_2}{\Lambda^2}-\frac{4\,\gamma}{\Lambda^2}\right)}\right).\label{D7} 
\end{eqnarray}

From the expressions in Eqs. (\ref{D6}) and (\ref{D7}), it is evident that the effective potential governing the spin-one vector perturbations is significantly influenced by the coupling constants present in the modified gravitational theories. In the context of $f(\mathcal{R})$-gravity, the effective potential is altered by the coupling constants $\alpha_2$, $\alpha_3$, and $\alpha_4$, which arise from the functional form of the gravitational action and the associated field equations. These coupling constants determine the strength and nature of the interaction between the vector field and the modified curvature terms, thereby affecting the propagation of scalar fields.

Similarly, in the framework of $\mathcal{RI}$-gravity, the effective potential is affected by a different set of coupling constants: $\alpha_2$, $\beta_1$, $\beta_2$, and $\gamma$. These constants arise from the specific modification used in $\mathcal{RI}$-gravity. The interplay between these parameters governs how the spin-one vector field interacts with the modified gravitational background, which affects the dynamics of scalar perturbations in this model.

\begin{figure}[ht!]
    \centering
    \includegraphics[width=0.45\linewidth]{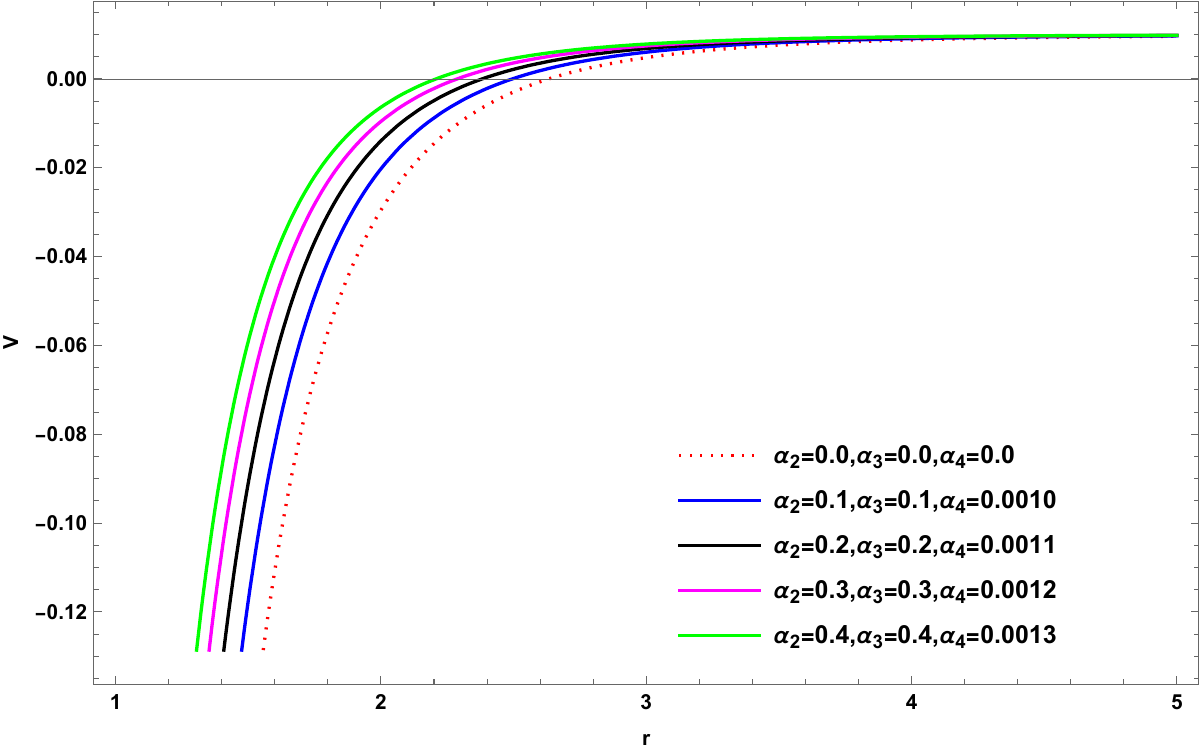}\quad\quad
    \includegraphics[width=0.45\linewidth]{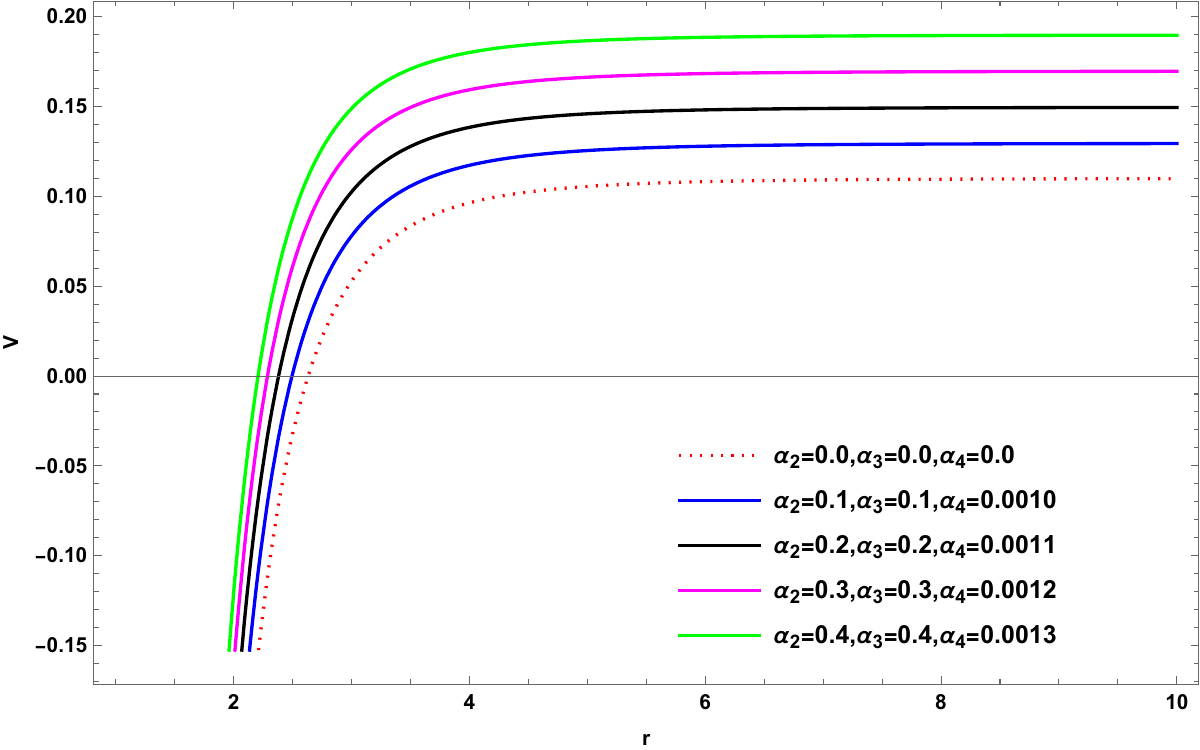}
    \caption{A comparison of the effective potential (\ref{D6}) with GR case for vector perturbations is plotted, considering cases: $m=0$ (left panel) and $m=1$ (right panel).  Here $M=1$, $\Lambda=-0.3$, and $k=0.1$.}
    \label{fig:3}
    \hfill\\
    \includegraphics[width=0.45\linewidth]{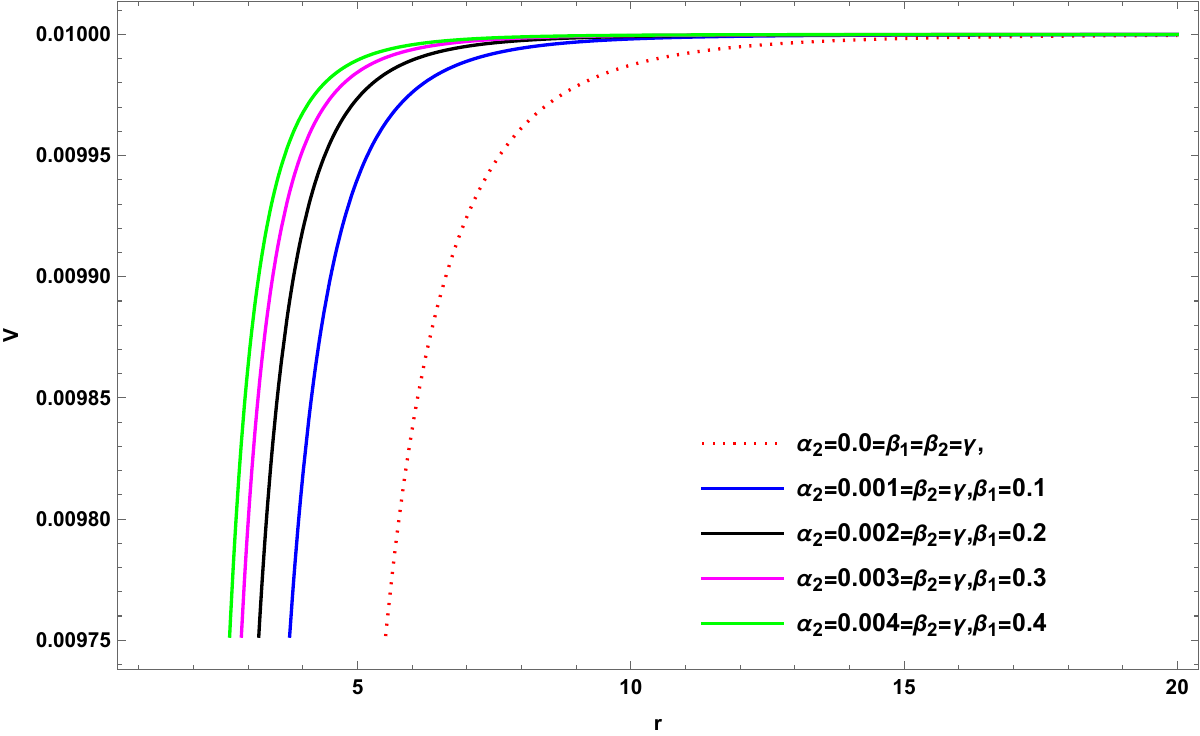}\quad\quad
    \includegraphics[width=0.45\linewidth]{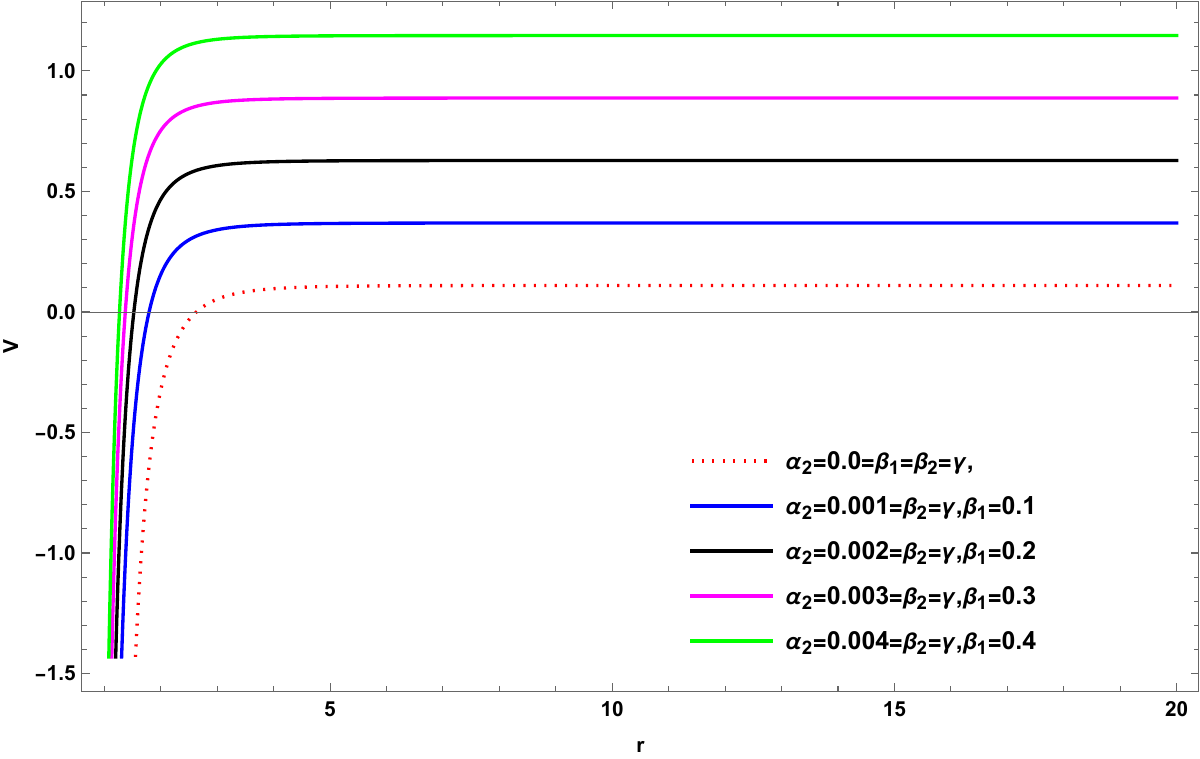}
    \caption{A comparison of the effective potential (\ref{D7}) with the GR case for vector perturbations is plotted, considering $m=0$ (left panel) and $m=1$ (right panel). Here $M=1$, $\Lambda=-0.3$, and $k=0.1$.}
    \label{fig:4}
\end{figure}

In Figure \ref{fig:3}, we illustrates the effective potential for electromagnetic perturbations within the framework of $f(\mathcal{R})$ gravity, considering various values of the coupling constants. This figure demonstrate how the perturbations potential alters with changes in the values of the coupling constants $\alpha_2$, $\alpha_3$, and $\alpha_4$. As these constants increase, the potential gradually rises, indicating that the strength of the scalar perturbations is affected by the modifications to the gravitational action.

Furthermore, we compare this potential to the standard GR case by setting $\alpha_2 = 0$, $\alpha_3 = 0$, and $\alpha_4 = 0$, effectively reducing $f(\mathcal{R})$ gravity to the GR limit. This comparison highlights the differences between the $f(\mathcal{R})$ gravity theory and GR. In the figure, the red dotted lines represent the perturbations potential in the GR case, while the solid colored lines correspond to the potential in the $f(\mathcal{R})$ gravity framework with non-zero coupling constants. 

Similarly, in Figure \ref{fig:4}, the electromagnetic perturbations potential within the framework of $\mathcal{RI}$-gravity, considering various values of the coupling constants is generated. The figure illustrates how the effective potential changes with variations in the coupling constants $\alpha_2$, $\beta_1$, $\beta_2$, and $\gamma$. As these constants increase, the potential gradually rises, suggesting that the strength and nature of the scalar perturbations are influenced by the modifications to the gravitational action introduced by these constants.

\section{Effect of GUP on Quantum Tunneling of Scalar Particles in $f(\mathcal{R})$ and $\mathcal{RI}$-gravity theories} \label{sec4}

Hawking radiation \cite{Hawking:1974rv,Gibbons:1977mu,Hawking:1982dh,Parikh:1999mf,Srednicki:1993im,Hawking:1976de,Callan:1992rs,Kanzi:2019gtu,Kanzi:2021cbg,Ovgun:2019ygw,Sakalli:2018nug,Al-Badawi:2020htj,Sakalli:2023pgn,Sucu:2023moz,Kanzi:2023itu}
 is a quantum mechanical phenomenon that arises near the event horizon of black holes. Modifications to the standard Hawking radiation, inspired by the Generalized Uncertainty Principle (GUP) \cite{Maggiore:1993rv,Scardigli:1999jh,Konishi:1989wk,Sakalli:2022swm}, provide a window into the interplay between quantum mechanics and gravity. The GUP-modified Klein-Gordon equation (GUPKGE) is a central tool for exploring these corrections \cite{Sakalli:2016mnk}. In this section, we calculate the GUP-modified Hawking radiation of cylindrical black holes described by two metrics in $f(R)$-gravity \eqref{CC10} and $\mathcal{RI}$-gravity \eqref{B11}. To this end, we employ the quantum tunneling approach and solve the GUP-modified wave equation for scalar fields propagating in these backgrounds. However, since both metrics of those modified gravity theories are structurally identical to each other, we consider Eq. \eqref{B10} as a generic line-element of cylindrical black hole of cosmological constant $\Lambda$. Then, we shall replace $\Lambda$ with $\Lambda^{f(\mathcal{R})}_m$ \eqref{CC9} and $\Lambda^{\mathcal{RI}}_m$ \eqref{B7}, respectively.

The GUPKGE for a scalar field $\Psi$ is given by \cite{Sakalli:2016mnk}
\begin{equation} \label{iz1}
- (i\,\hbar)^2 \partial^2_t \Psi=\left[(i\,\hbar)^2\, \partial^i\,\partial_i + m_p^2 \right] \left[1 - 2\,\alpha_{\text{GUP}}\, \left((i\,\hbar)^2 \partial^i\, \partial_i + m_p^2 \right) \right]\,\Psi,
\end{equation}
where $\alpha_{\text{GUP}}$ is the GUP parameter and $m_p$ is the mass of the scalar particle. The above equation includes quantum gravitational corrections due to the GUP and will be used to derive the corrected Hawking temperature.

To solve the GUPKGE \eqref{iz1}, we assume the following wave function for the scalar field:
\begin{equation} \label{iz2}
\Psi(t, r, \varphi, z) = \exp\left(\frac{i}{\hbar} \mathcal{S}(t, r, \varphi, z)\right),
\end{equation}
where $\mathcal{S}(t, r, \varphi, z)$ is the action, expressed as:
\begin{equation} \label{iz3}
\mathcal{S}(t, r, \varphi, z) = -E\, t + W(r) + j\, \varphi + k\, z + K.
\end{equation}
Here, $E$ is the energy, $j$ represents the angular momentum, $k$ denotes the separation constant associated with the $z$-direction, and $K$ is a complex constant. Substituting $\Psi$ into the KGE and retaining only the leading-order terms in $\hbar$, we obtain:
\begin{equation} \label{iz4}
\frac{1}{g_{tt}} (\partial_t \mathcal{S})^2 = g_{rr} (\partial_r \mathcal{S})^2 + \frac{j^2}{g_{\varphi\varphi}} + \frac{k^2}{g_{zz}} + m_p^2 \left[ 1 - 2\alpha_\text{GUP} \left( g_{rr} (\partial_r \mathcal{S})^2 + \frac{j^2}{g_{\varphi\varphi}} + \frac{k^2}{g_{zz}} + m_p^2 \right) \right].
\end{equation}

After substituting the ansatz \eqref{iz3} in Eq. \eqref{iz4} and focusing only on radial motion ($j = 0$, $k = 0$) and substituting the equation can be simplified to
\begin{equation} \label{iz5}
\frac{1}{g_{tt}} E^2 = g_{rr} (W'(r))^2 + m_p^2 \left[ 1 - 2\alpha_\text{GUP} \left( g_{rr} (W'(r))^2 + m_p^2 \right) \right].
\end{equation}

Reorganizing and solving for $W'(r)$, we get:
\begin{equation} \label{iz6}
W'(r) = \pm \sqrt{\frac{\frac{E^2}{g_{tt}} - m_p^2 \left( 1 - 2\alpha_\text{GUP} m_p^2 \right)}{g_{rr}\left(1 - 2\alpha_\text{GUP} m_p^2\right)}}.
\end{equation}
which naturally yields the following integral solution for $W(r)$:
\begin{equation} \label{iz7}
W(r) = \pm \int \sqrt{\frac{\frac{E^2}{g_{tt}} - m_p^2 \left( 1 - 2\alpha_\text{GUP} m_p^2 \right)}{g_{rr}\left(1 - 2\alpha_\text{GUP} m_p^2\right)}}
\end{equation}

Close to the horizon \( r_h \), where \( g_{tt}(r_h) = 0 \), the metric function \( \Delta(r) = g_{tt}(r) \) can be expanded as:
\begin{equation} \label{iz8}
\Delta(r) \approx \Delta'(r_h) (r - r_h)=2\kappa(r-r_h),
\end{equation}
where $\Delta'(r_h)$ is the derivative of $\Delta(r)$ with respect to $r$ at the event horizon and $\kappa$ denotes the surface gravity: 
\begin{equation} \label{iz9}
\Delta'(r_h) =2\kappa= -\frac{2\Lambda}{3}\,r_h + \frac{4\,M}{\sqrt{-\frac{\Lambda}{3}}\,r_h^2}.
\end{equation}
Moreover, the radius of the event horizon is expressed as follows:
\begin{equation} \label{iz10}
r_h=\sqrt{-\frac{3}{\Lambda}}\,(4\,M)^{1/3}.
\end{equation}
Using Eq. \eqref{iz8} and the metric functions, the integral for $W(r)$ near the horizon simplifies to:
\begin{equation} \label{iz11}
W(r) = \pm \frac{1}{\sqrt{1 - 2m_p^2 \alpha_\text{GUP}}} \int \frac{\sqrt{E^2 - \Delta'(r_h) (r - r_h) m_p^2 (1 - 2m_p^2 \alpha_\text{GUP})}}{\sqrt{\Delta'(r_h) (r - r_h)}} \, dr.
\end{equation}

Let $ x = r - r_h $, so that $\Delta(r) \approx \Delta'(r_h) x $. The integral becomes:
\begin{equation} \label{iz12}
W(r) = \pm \frac{1}{\sqrt{1 - 2m_p^2 \alpha_\text{GUP}}} \frac{1}{\sqrt{\Delta'(r_h)}} \int \sqrt{\frac{E^2}{x} - m_p^2 (1 - 2m_p^2 \alpha_\text{GUP})} \, dx.
\end{equation}

For small $ x $ near the pole, i.e., around the event horizon, $\frac{E^2}{x}$  dominates, so:
\begin{equation} \label{iz13}
W(r) \approx \pm \frac{1}{\sqrt{1 - 2m_p^2 \alpha_\text{GUP}}} \frac{1}{\sqrt{\Delta'(r_h)}} \int \frac{E}{\sqrt{x}} \, dx.
\end{equation}

Since the integral of \( \frac{1}{\sqrt{x}} \) is trivial:
\begin{equation} \label{iz14}
\int \frac{1}{\sqrt{x}} \, dx = 2\sqrt{x},
\end{equation}
thus, we get
\begin{equation} \label{iz15}
W(r) \approx \pm \frac{2E}{\sqrt{\left(1 - 2m_p^2 \alpha_\text{GUP}\right)\Delta'(r_h)}} \sqrt{r - r_h}.
\end{equation}

To compute the imaginary part around the pole, we employ the Feyman's prescription method \cite{Wald:1979kp,Visser:2021ucg}
 and performing some manipulations, we obtain
\begin{equation} \label{iz16}
\text{Im}(W(r)) = \pm \frac{\pi E}{\sqrt{\Delta'(r_h) (1 - 2m_p^2 \alpha_\text{GUP})}}.
\end{equation}
The designation of a negative (positive) sign denotes an ingoing (outgoing) bosons. It is pertinent to acknowledge that the well-known issue of the factor two discrepancy present in the aforementioned expression, which results in an incorrect estimation of the tunneling rate, can be rectified through a method delineated in \cite{Akhmedova:2008dz,Akhmedova:2008au}
. Alternatively, this issue can be addressed by assigning a probability of $100 \%$ to the scenario involving incoming bosons. Specifically, we express this as
\begin{equation} \label{iz17}
\mathcal{P}_{-} \simeq e^{-2 \text{Im} W_{-}}=1,
\end{equation}
which consequently results in
\begin{equation} \label{iz18}
\text{Im} \mathcal{S}_{-}=\text{Im} W_{-}+K=0    
\end{equation}
Conversely, when considering the boson being emitted, we observe that
\begin{equation} \label{iz19}
\text{Im} \mathcal{S}_{+}=\text{Im} W_{+}+K.
\end{equation}
Based on Eq. \eqref{iz16}, it becomes evident that $W_{+}=-W_{-}$. Consequently, the tunneling probability associated with the emitted bosons can be interpreted in the following manner
\begin{equation} \label{iz20}
 \mathcal{P}_{+}=e^{-2  \text{Im} \mathcal{S}_{+}} \simeq e^{-4 \operatorname{Im} W_{+}}    
\end{equation}
Finally, using Eqs. \eqref{iz17} and \eqref{iz20} the tunneling rate of bosons becomes
\begin{equation} \label{iz21}
 \mathcal{T_{R}}=\frac{\mathcal{P}_{+}}{\mathcal{P}_{-}} \simeq e^{\left(-4 \text{Im} W_{+}\right)} .   
\end{equation}
Comparison of Eq. \eqref{iz21} with the Boltzmann factor \cite{Parikh:1999mf} $\left(\mathcal{T_{R}}=exp(-\frac{E}{T})\right)$, we derive the GUP-corrected Hawking temperature as follows
\begin{equation} \label{iz22}
T_\text{GUP} = T_H\sqrt{1 - 2m_p^2 \alpha_\text{GUP}}.
\end{equation}
where the standard Hawking temperature is given by \cite{Wald:1984rg}
\begin{equation} \label{iz23}
T_H = \frac{\kappa}{2\pi}=\frac{\sqrt{\Delta'(r_h)}}{4\pi}.
\end{equation}
Substituting $\Delta'(r_h) =\sqrt{-3\,\Lambda}\,(4\,M)^{1/3}$, the explicit form of $T_H$ reads
\begin{equation} \label{iz24}
T_H =\frac{\sqrt{\sqrt{-3\,\Lambda}\,(4\,M)^{1/3}}}{4\pi}.
\end{equation}
Therefore, the GUP-corrected Hawking temperatures of the cylindrical black holes in $f(\mathcal{R})$ and $\mathcal{RI}$-gravity theories are given by
\begin{align}\label{iz25}
  T^{f(\mathcal{R})}_\text{GUP}& = T^{f(\mathcal{R})}_H\sqrt{1 - 2m_p^2 \alpha_\text{GUP}},\\
    T^{\mathcal{RI}}_\text{GUP} &= T^{\mathcal{RI}}_H\sqrt{1 - 2m_p^2 \alpha_\text{GUP}}, \label{iz26}
      \end{align}
in which the unmodified Hawking radiations are as follows
\begin{align} \label{iz27n}
T^{f(\mathcal{R})}_H&=\frac{\sqrt{\sqrt{-3\,\Lambda^{f(\mathcal{R})}_m}\,(4\,M)^{1/3}}}{4\pi}=\frac{\sqrt{\sqrt{3\,(-\Lambda+16\,\Lambda^3\,\alpha_2+128\,\Lambda^4\,\alpha_3+768\,\alpha_4\,\Lambda^5)}\,(4\,M)^{1/3}}}{4\pi}, \\
T^{\mathcal{RI}}_H&=\frac{\sqrt{\sqrt{-3\,\Lambda^{\mathcal{RI}}_m}\,(4\,M)^{1/3}}}{4\pi}=\frac{\sqrt{\sqrt{3\,\left(-\Lambda+16\,\Lambda^3\,\alpha_2-\frac{3\,\beta_1}{\Lambda}-\frac{16\,\beta_2}{\Lambda^2}-\frac{4\,\gamma}{\Lambda^2}\right)}\,(4\,M)^{1/3}}}{4\pi}. \label{iz28n}
\end{align}

\begin{figure}[ht!]
    \centering
    \includegraphics[scale=0.6]{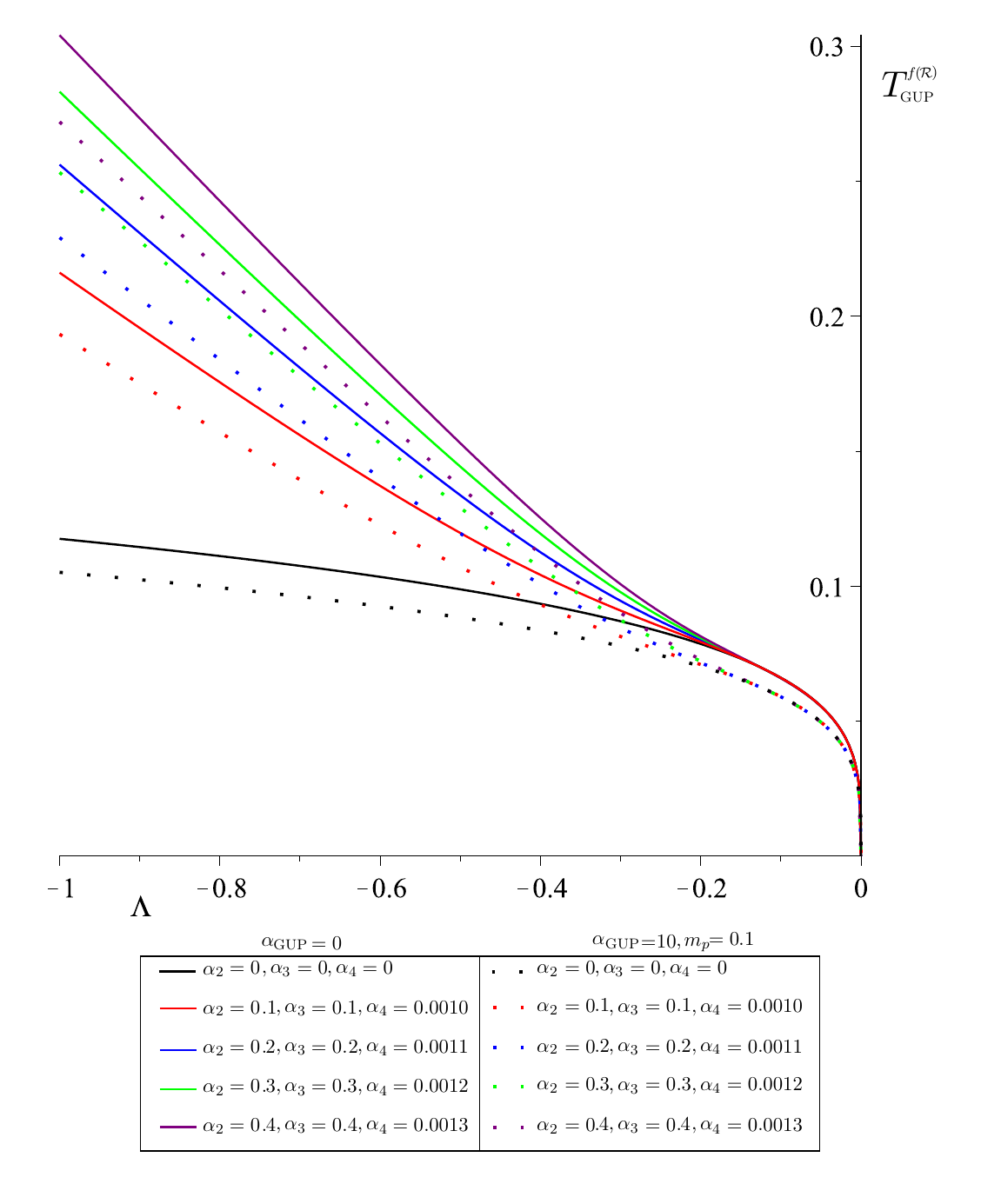}\quad\quad
    \caption{Comparison of the GUP-modified Hawking temperatures for cylindrical black holes in $f(R)$-gravity under varying coupling constants $\alpha_2$, $\alpha_3$, $\alpha_4$. The results highlight the quantum corrections to black hole evaporation rates induced by the GUP parameter $\alpha_{\mathrm{GUP}}$ and the scalar particle mass $m_p$. Variations in $\alpha_2$, $\alpha_3$, $\alpha_4$ showcase significant deviations in Hawking temperature compared to GR (black solid line), emphasizing the impact of higher-order curvature terms in $f(R)$-gravity. The plots are governed by Eqs. \eqref{iz27n} and \eqref{iz28n}. The mass parameters  are chosen as $M=1$ and $m_p=0.1$.}

    \label{fig:iz1}
        \hfill\\
\end{figure}

\begin{figure}[ht!]
    \centering
    \includegraphics[scale=0.6]{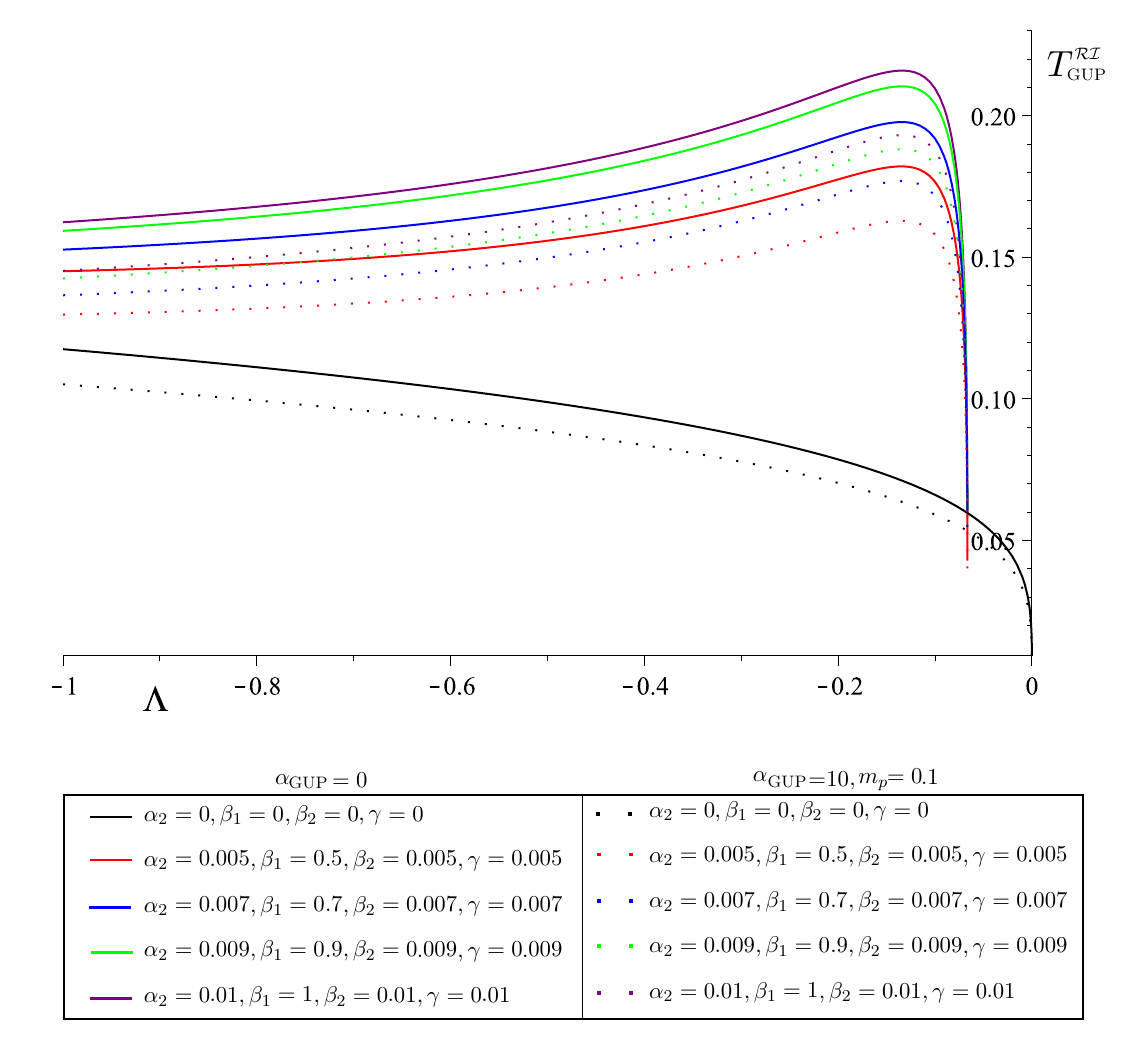}\quad\quad
    \caption{The effects of the GUP parameter $\alpha_{\mathrm{GUP}}$ and coupling constants $\alpha_2$, $\beta_1$, $\beta_2$, and $\gamma$ on the modified Hawking temperature $T_{\text{GUP}}^{\mathcal{RI}}$ for cylindrical black holes  in $\mathcal{RI}$-gravity. The results are shown for two cases: $\alpha_{\mathrm{GUP}} = 0$ (classical scenario) and $\alpha_{\mathrm{GUP}} = 10$ (quantum gravity corrections). In GR limit, the $\mathcal{RI})$-gravity parameters are set to zero, the Hawking temperature goes zero as $\Lambda$ approaches zero. However, when $\mathcal{RI})$-gravity is activated, the Hawking temperature remains finite as $\Lambda$ tends to zero, signaling the possibility of remnant formation.  Variations in $\alpha_2$, $\beta_1$, $\beta_2$, and $\gamma$ reveal significant deviations from GR, emphasizing the role of $\mathcal{RI})$-gravity corrections. The plots are governed by Eqs. \eqref{iz27n} and \eqref{iz28n}. The mass parameters  are chosen as $M=1$ and $m_p=0.1$.}

    \label{fig:iz2}
        \hfill\\
\end{figure}

Figure \ref{fig:iz1} demonstrates the influence of the GUP parameter $\alpha_{\mathrm{GUP}}$ and the coupling constants $\alpha_2$, $\alpha_3$, and $\alpha_4$ on the modified Hawking temperature $T_{\text{GUP}}^{f(R)}$ for cylindrical black holes in $f(R)$-gravity. The plot includes two scenarios: $\alpha_{\mathrm{GUP}} = 0$ (no GUP corrections) and $\alpha_{\mathrm{GUP}} = 10, m_p = 0.1$ (with GUP corrections), allowing for a direct comparison of quantum gravity effects. The Hawking temperature decreases as the cosmological constant $\Lambda$ approaches $0$, highlighting the suppressive effect of $\alpha_{\mathrm{GUP}}$ on the black hole evaporation rate. Higher values of $\alpha_2$, $\alpha_3$, and $\alpha_4$ further amplify the deviations in $T_{\text{GUP}}^{f(R)}$ from GR, underscoring the significant role of higher-order curvature corrections in $f(R)$-gravity. On the other hand, Fig. \ref{fig:iz2} illustrates the distinct thermodynamic behavior of cylindrical black holes in $\mathcal{RI}$-gravity under the influence of the GUP parameter $\alpha_{\mathrm{GUP}}$ and coupling constants $\alpha_2$, $\beta_1$, $\beta_2$, and $\gamma$. In the GR limit, where $\alpha_{\mathrm{GUP}}$ and RI parameters are all set to zero, the Hawking temperature $T_{\text{GUP}}^{\mathcal{RI}}$ vanishes as $\Lambda \to 0$, indicating complete black hole evaporation. However, when $\mathcal{RI}$-gravity is switched on, the Hawking temperature remains finite even as $\Lambda$ approaches zero, strongly suggesting the formation of a black hole remnant \cite{Aydemir:2020pao,Fernandes:2021dsb}, underscores the non-trivial contributions of the $\mathcal{RI}$-gravity parameters to black hole thermodynamics. Additionally, the inclusion of GUP corrections further modifies the evaporation profile, highlighting the interplay between quantum gravity effects and the modifications introduced by $\mathcal{RI}$-gravity . The observed deviations from GR are amplified by higher-order curvature corrections governed by $\alpha_2$, $\beta_1$, $\beta_2$, and $\gamma$. 

In summary, the analysis of GUP-modified Hawking temperatures for cylindrical black holes in $f(\mathcal{R})$-gravity and $\mathcal{RI}$-gravity reveals significant quantum gravitational corrections. Specifically, the coupling constants $\alpha_2$, $\alpha_3$, and $\alpha_4$ in $f(\mathcal{R})$-gravity theory significantly alters the GUP-corrected Hawking temperature, modifying the black hole evaporation rate compared to that in GR. Similarly, in the context of $\mathcal{RI}$-gravity theory, the coupling constants $\alpha_2$, $\beta_1$, $\beta_2$, and $\gamma$ were shown to significantly affect the GUP-corrected temperature. Moreover, the GUP introduces a correction factor $\sqrt{1 - 2\,m_p^2\, \alpha_\text{GUP}}$ that reduces the black hole's temperature, slowing its evaporation rate. This modification depends on the GUP parameter $\alpha_\text{GUP}$ and the scalar particle mass $m_p$, emphasizing the quantum mechanical influence on black hole thermodynamics. While the structural similarity of the two gravity theories results in analogous expressions for GUP-corrected temperatures, the unmodified Hawking temperatures differ due to their dependence on the respective cosmological constants, $\Lambda_m^{f(\mathcal{R})}$ and $\Lambda_m^{\mathcal{RI}}$. These findings highlight the potential for testing quantum gravitational theories through observations of black hole radiation spectra, bridging the gap between theoretical predictions and experimental constraints.

Finally, we want to represent the GUP-modified entropy \cite{Medved:2004yu,Sakalli:2022swm,Khosropour:2024boy,Chen:2024mlr,Sucu:2024gtr}
 of black holes, which can be derived by integrating the first law of black hole thermodynamics \cite{Wald:1984rg}: 
\begin{equation} \label{iz27}
dS = \frac{dM}{T_\text{GUP}}, 
\end{equation}
where $T_\text{GUP}$ is the GUP-modified Hawking temperature obtained earlier. Substituting $T_\text{GUP}$ and using the standard Hawking temperatures $T_H$, the entropy is given by:
\begin{equation} \label{iz27b}
S_\text{GUP} = \int \frac{dM}{T_H \sqrt{1 - 2\,m_p^2\, \alpha_\text{GUP}}}.
\end{equation}

Using the relation $dM = \kappa\, dA / (8\,\pi)$, where $A$ is the area of the horizon and $\kappa$ is the surface gravity, and substituting $T_H$ in terms of $\kappa$, we can rewrite the entropy expression as:
\begin{equation} \label{iz28}
S_\text{GUP} = \frac{1}{4} \int \frac{dA}{\sqrt{1 - 2\,m_p^2\, \alpha_\text{GUP}}}.
\end{equation}

Factoring out the correction term $\sqrt{1 - 2\,m_p^2\, \alpha_\text{GUP}}$, the integration becomes:
\begin{equation} \label{iz29}
S_\text{GUP} = \frac{1}{4\sqrt{1 - 2\,m_p^2\, \alpha_\text{GUP}}} \int dA.
\end{equation}

Since the area of the horizon $A$ is proportional to $r_h^2$ (the radius of the event horizon squared), $dA$ can be expressed in terms of $r_h$:
\begin{equation} \label{iz30}
A = 2\,\pi\, z\, r_h, \quad\quad dA = 2\,\pi\, z\,  dr_h.
\end{equation}

Thus, the entropy becomes:
\begin{equation} \label{iz31}
S_\text{GUP} = \frac{1}{4\sqrt{1 - 2\,m_p^2\, \alpha_\text{GUP}}} \int 2\,\pi\, z\, r_h\, dr_h.
\end{equation}

Performing the integration yields:
\begin{equation} \label{iz32}
S_\text{GUP} = \frac{\pi\, z}{4\sqrt{1 - 2\,m_p^2\, \alpha_\text{GUP}}}\, r_h^2 + C,
\end{equation}
where $C$ is the integration constant. By setting $C = 0$ for consistency with the standard Bekenstein-Hawking entropy \cite{Sachdev:2015efa} in the absence of GUP corrections, we obtain:
\begin{equation} \label{iz33}
S_\text{GUP} = \frac{\pi\, z r_h^2}{4\,\sqrt{1 - 2\,m_p^2\, \alpha_\text{GUP}}}.
\end{equation}

Substituting the horizon radius $r_h =\sqrt{-\frac{3}{\Lambda}}\,(4\,M)^{1/3}$, the GUP-modified entropy can be explicitly expressed as:
\begin{equation} \label{iz34}
S_\text{GUP} = \frac{\pi\, z}{\sqrt{1 - 2\,m_p^2\, \alpha_\text{GUP}}}\frac{(27M^2/4)^{1/3}}{(-\Lambda)}.
\end{equation}

Therefore, in the context of modified gravity theories, as discussed earlier, we find the GUP-modified entropy as follows:
\begin{align}
S^{f(\mathcal{R})}_\text{GUP}& = \frac{\pi\, z}{\sqrt{1 - 2\,m_p^2\, \alpha_\text{GUP}}}\,\frac{(27M^2/4)^{1/3}}{\left(-\Lambda+16\,\Lambda^3\,\alpha_2+128\,\Lambda^4\,\alpha_3+768\,\alpha_4\,\Lambda^5\right)}, \label{iz35}\\
S^{\mathcal{RI}}_\text{GUP}& = \frac{\pi\, z}{\sqrt{1 - 2\,m_p^2\, \alpha_\text{GUP}}}\,\frac{(27M^2/4)^{1/3}}{\left(-\Lambda+16\,\Lambda^3\,\alpha_2-\frac{3\,\beta_1}{\Lambda}-\frac{16\,\beta_2}{\Lambda^2}-\frac{4\,\gamma}{\Lambda^2}\right)}.\label{iz36}
\end{align}

\begin{figure}[ht!]
    \centering
    \includegraphics[scale=0.6]{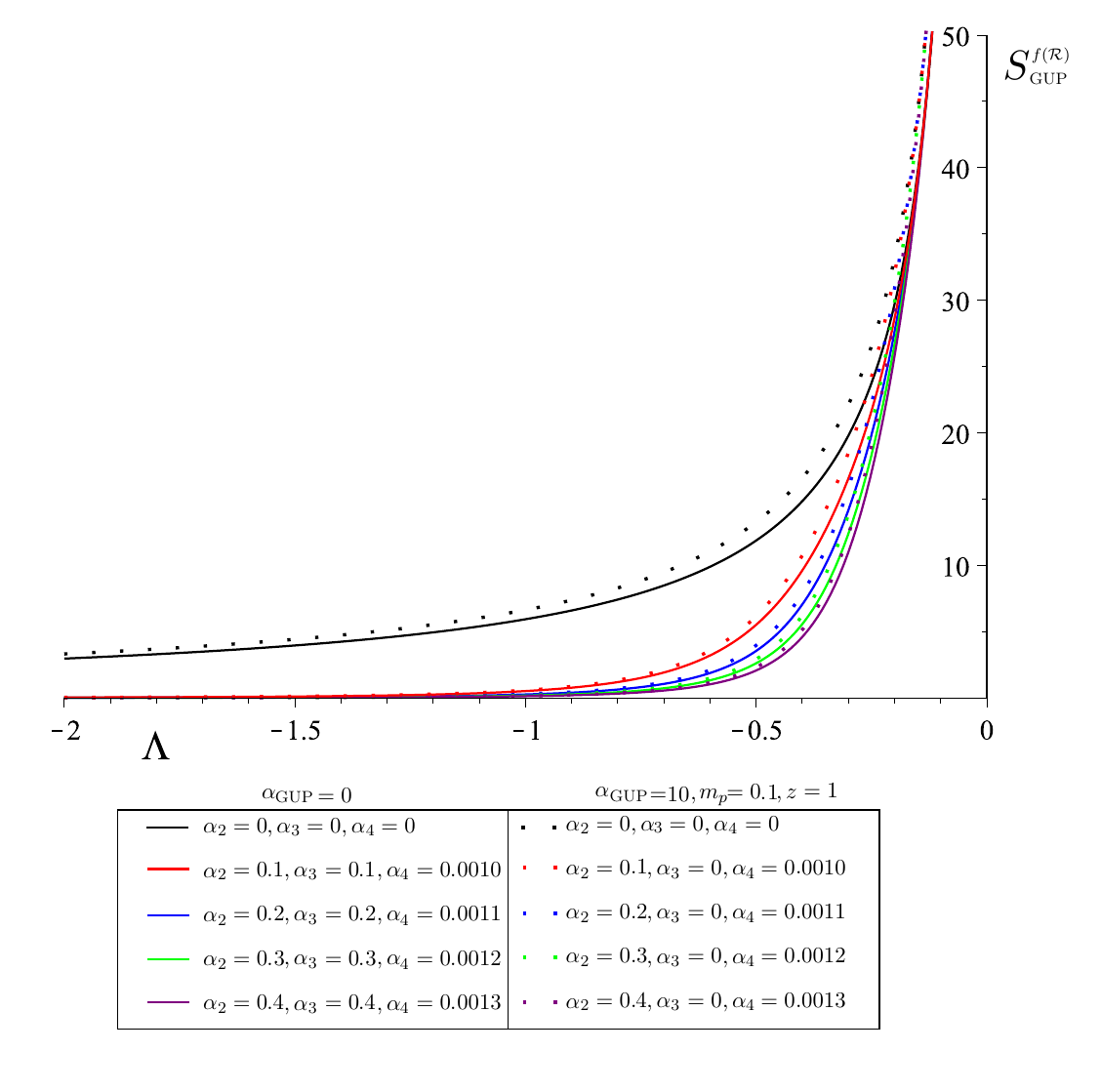}\quad\quad
    \caption{Behavior of the GUP-modified entropy $S_{\text{GUP}}^{f(\mathcal{R})}$ for cylindrical black holes in $f(\mathcal{R})$-gravity as a function of the cosmological constant $\Lambda$. The results are plotted for two scenarios: $\alpha_{\mathrm{GUP}} = 0$ (classical limit) and $\alpha_{\mathrm{GUP}} = 10$ (quantum gravity corrections). The entropy increases rapidly as $\Lambda$ approaches zero, with significant differences observed between the classical and quantum gravity cases. Higher values of the coupling constants $\alpha_2$, $\alpha_3$, and $\alpha_4$ enhance the deviations from the GR limit. The plots are governed by Eq. \eqref{iz35}. The physical parameters  are chosen as $M=1$, $m_p=0.1$, and $z=1$. }
    \label{fig:iz3}
\end{figure}

\begin{figure}[ht!]
    \centering
    \includegraphics[scale=0.6]{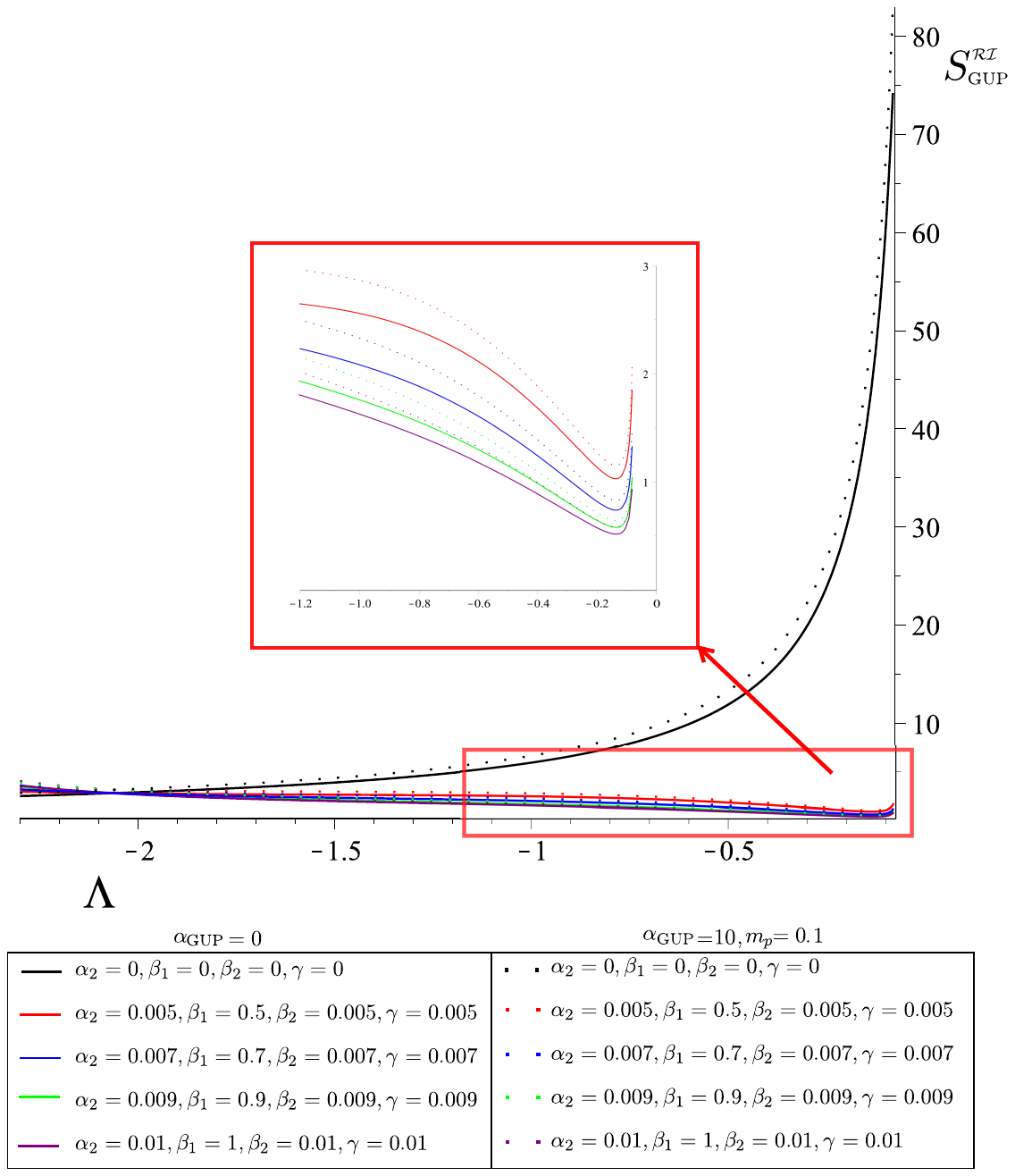}\quad\quad
    \caption{GUP-modified entropy $S_{\text{GUP}}^{\mathcal{RI}}$ for cylindrical black holes in $\mathcal{RI}$-gravity as a function of the cosmological constant $\Lambda$. The plot highlights two cases: $\alpha_{\mathrm{GUP}} = 0$ (classical limit) and $\alpha_{\mathrm{GUP}} = 10$ (quantum gravity corrections). The inset zooms in on the low-entropy region near $\Lambda = -0.5$, showcasing subtle variations induced by the coupling constants $\alpha_2$, $\beta_1$, $\beta_2$, and $\gamma$. As $\Lambda$ approaches zero, the entropy tends to increase, with larger deviations observed for lower values of the coupling constants. The plots are governed by Eq. \eqref{iz36}. The physical parameters  are chosen as $M=1$, $m_p=0.1$, and $z=1$. }
    \label{fig:iz4}
\end{figure}

The GUP-modified entropy (\ref{iz34}) of the cylindrical black hole demonstrate a reduction relative to the classical Bekenstein-Hawking entropy due to the factor $\sqrt{1-2\,m_p^2\,\alpha_\text{GUP}}$. This modification, which relies on the GUP parameter $\alpha_\text{GUP}$ as well as the mass of the scalar particle denoted by $m_p$, highlights the significant role that quantum gravitational phenomena play in affecting the thermodynamic behavior of black holes.

By setting $\alpha_2 = 0$, $\alpha_3 = 0$, and $\alpha_4 = 0$, we reduce $f(\mathcal{R})$ gravity to the GR limit. This comparison underscores the differences between the modified $f(\mathcal{R})$ gravity theory and GR, offering insights into how the higher-order scalar curvature ($\mathcal{R}$) modifies the nature of the GUP-modified entropy (\ref{iz35}) compared to the results known in GR. Similarly, by setting $\alpha_2 = 0$, $\beta_1 = 0$, $\beta_2 = 0$, and $\gamma = 0$, we reduce $\mathcal{RI}$ gravity to the GR limit. This distinction highlights how the higher-order scalar curvature $(\mathcal{R})$, the anti-curvature tensor $(A^{\mu\nu}=R^{-1}_{\mu\nu})$ and its scalar $(\mathcal{A}=g_{\mu\nu}\,A^{\mu\nu})$, influence the GUP-modified entropy (\ref{iz36}), offering a clear contrast to the results in GR. Figures \ref{fig:iz3} and \ref{fig:iz4} provide key insights into the thermodynamic behavior of cylindrical black holes in $\mathcal{RI}$-gravity under the influence of GUP modifications. In the GR limit, where the $\mathcal{RI}$ parameters are all set to zero, the entropy $S_{\text{GUP}}^{f(R)}$ makes a sudden icrease as $\Lambda \to 0$. This behavior is indicative of a complete evaporation scenario typical of classical black hole thermodynamics. However, when $\mathcal{RI}$-gravity is switched on, a starkly different trend is observed: the entropy first decreases with $\Lambda$ in the low-cosmological constant regime, but it never reaches zero as $\Lambda$ approaches zero. This suggests the presence of a residual entropy, potentially signaling the formation of black hole remnants.

The possibility of remnant formation is consistent with theoretical predictions in the literature \cite{Aydemir:2020pao, Fernandes:2021dsb}, where the inclusion of modified gravity corrections prevents the complete evaporation of black holes. In particular, $\mathcal{RI}$-gravity introduces a stabilizing effect on the black hole's thermodynamic evolution, supported by the GUP-induced quantum gravity corrections. Figure \ref{fig:iz3} highlights how higher-order coupling constants $\alpha_2$, $\alpha_3$, and $\alpha_4$ further amplify deviations from the GR predictions, while Fig. \ref{fig:iz4} emphasizes the interplay between GUP corrections and $\mathcal{RI}$ parameters. Notably, the inset in Figure 8 reveals that entropy remains finite even in regions where classical entropy would diverge, underscoring the non-trivial effects of $\mathcal{RI}$-gravity on the black hole's thermodynamic end state.

\section{Conclusions}\label{sec5}

This study provided a comprehensive analysis of the scalar perturbations and the GUP-modified thermodynamic properties of cylindrical black holes constructed within the context of modified gravity theories, specifically $f(\mathcal{R})$-gravity and $\mathcal{RI}$-gravity theories. We began by re-deriving the cylindrical black hole solutions within these modified gravity theories. The effective cosmological constants were shown to depend on coupling constants $\alpha_i$, $\beta_i$, and $\gamma$, which characterize the respective theories, as reported in \cite{arxiv}. The results revealed that these coupling constants significantly influenced the black hole dynamics compared to General Relativity (GR).

Using the Klein-Gordon equation \eqref{C1}, we analyzed a massless scalar field in the background of the black hole constructed within the framework of $f(\mathcal{R})$-gravity and $\mathcal{RI}$-gravity theories. In this context, we derived the effective potential $V$, as expressed in Eq. \eqref{C8}. We showed that this effective potential for scalar perturbations was influenced by the constants $\alpha_2$ coupled with $\mathcal{R}^2$, $\alpha_3$ coupled with $\mathcal{R}^3$, and $\alpha_4$ coupled with $\mathcal{R}^4$ in $f(\mathcal{R})$-gravity theory. Similarly, in $\mathcal{RI}$-gravity theory, the scalar perturbation potential was influenced by the constants $\alpha_2$ coupled with $\mathcal{R}^2$, $\beta_1$ coupled with $\mathcal{A}$, $\beta_2$ coupled with $\mathcal{A}^2$, and $\gamma$ coupled with $A^{\mu\nu}A_{\mu\nu}$. Figures \ref{fig:1}–\ref{fig:2} illustrated the behavior of the effective potential $V(r)$ in both $f(\mathcal{R})$-gravity and $\mathcal{RI}$-gravity theories, and compared it with the results in the GR limit.

Moreover, Hawking radiation was analyzed within the framework of the Generalized Uncertainty Principle (GUP). The GUP-corrected Hawking temperatures for cylindrical black holes constructed in the context of $f(\mathcal{R})$-gravity and $\mathcal{RI}$-gravity theories were derived, as presented in Eqs.~\eqref{iz25} and \eqref{iz26}, which highlighted the quantum corrections to black hole evaporation. The analysis demonstrated that the GUP parameter $\alpha_\text{GUP}$ causes a reduction in the Hawking temperature, leading to a slower rate of black hole evaporation compared to the standard result in General Relativity (GR). This slower evaporation is a direct consequence of the modified dispersion relations that arise from the GUP, which in turn affect the thermodynamic properties of the black holes.

Additionally, the coupling constants of the modified gravity theories were shown to have significant effects on the GUP-corrected Hawking temperature. Specifically, the coupling constants $\alpha_2$, $\alpha_3$, and $\alpha_4$ in $f(\mathcal{R})$-gravity theory were found to influence the GUP-corrected temperature, modifying the black hole evaporation rate compared to that in GR. Similarly, in the context of $\mathcal{RI}$-gravity theory, the coupling constants $\alpha_2$, $\beta_1$, $\beta_2$, and $\gamma$ were shown to significantly affect the GUP-corrected temperature. These modifications resulted in notable deviations from the GUP-corrected temperature in the GR limit, thus emphasizing the impact of the modified gravity framework on the thermodynamic properties of the black holes.

Furthermore, the GUP-corrected entropy was also calculated, as expressed in Eq.~\eqref{iz34}, which indicated a significant deviation from the classical Bekenstein-Hawking entropy. This deviation arises due to the quantum modifications introduced by the GUP, which alter the black hole's entropy at the microscopic level. We also derived the entropy expressions (\ref{iz35}) and (\ref{iz36}), respectively for cylindrical black holes in both $f(\mathcal{R})$-gravity and $\mathcal{RI}$-gravity theories, highlighting the role of the coupling constants in these modified theories. In both cases, we observed that the coupling constants influenced the GUP-corrected entropy, leading to substantial changes compared to the classical entropy expression in General Relativity. These results underscored the importance of considering both the GUP and modified gravity theories when analyzing the thermodynamic properties of black holes, as the coupling constants played a crucial role in modifying the entropy and temperature, thus providing deeper insights into the quantum nature of black hole thermodynamics.

Future directions for this study involve several promising extensions that could deepen our understanding of black hole thermodynamics within modified gravity frameworks. One key direction is to extend the analysis to rotating black holes (Kerr-like solutions) and higher-dimensional black holes within these modified gravity theories. The inclusion of rotation and additional spatial dimensions could provide richer dynamics, such as the effects of angular momentum on the GUP-modified thermodynamic properties, Hawking radiation, and the scalar perturbations, potentially unveiling new phenomena not present in the static, four-dimensional case.

Another crucial avenue for future research is the exploration of the observational implications of the modified Hawking radiation spectrum. Since Hawking radiation is theorized to be extremely weak and difficult to detect, modifications introduced by the GUP and modified gravity could have significant effects on the emission characteristics, such as its temperature, spectrum, or time evolution. Investigating the potential for detecting these modified radiation signatures through black hole shadow imaging-such as that achieved by the Event Horizon Telescope-or through gravitational wave signals from black hole mergers could offer valuable experimental evidence supporting or challenging the theoretical predictions of GUP-modified black hole thermodynamics. These observations could also help differentiate between different modified gravity theories and provide deeper insights into the quantum nature of black holes.

Finally, an exciting direction would be to incorporate other quantum gravity approaches, such as loop quantum gravity (LQG), into this framework. LQG, which provides a non-perturbative quantization of spacetime, might introduce further modifications to black hole thermodynamics at both the classical and quantum levels. Exploring how these quantum gravity effects interplay with modified gravity theories could reveal new aspects of black hole structure, such as the nature of singularities, the role of spacetime discreteness, and the potential for a finite entropy or temperature at the event horizon. Integrating LQG with the GUP framework may uncover additional corrections to the Hawking radiation, entropy, and perturbation spectra, and could offer a unified view of the quantum structure of black holes across various quantum gravity paradigms.

These future directions promise to expand the scope of our current understanding and may pave the way for groundbreaking theoretical and observational advances in the study of black holes, quantum gravity, and the fundamental nature of spacetime.

\section*{Acknowledgments}

F.A. acknowledges the Inter University Centre for Astronomy and Astrophysics (IUCAA), Pune, India for granting visiting associateship.  \.{I}.~S. expresses gratitude to T\"{U}B\.{I}TAK, ANKOS, and SCOAP3 for their financial support. He also acknowledges COST Actions CA22113 and CA21106 for their contributions to networking.

\section*{Data Availability Statement}

This manuscript has no associated data. [Authors’ comment: This is a theoretical study and no experimental data has been listed.]

\section*{Code Availability Statement}

This manuscript has no associated code/software. [Authors’ comment: Code/software sharing not applicable to this article as no code/software was generated or analyzed in the current study.]

\section*{Conflict of interest}

Author(s) declare no such conflict of interest.


\begin{thebibliography}{99}


\bibitem{b1} S. Chandrasekhar, { \tt The Mathematical Theory of Black Holes}, Oxford University Press, Oxford (1992).

\bibitem{b2} B. Carr, K. Kohri, Y. Sendouda, J. Yokoyama, Phys. Rev. {\bf D 81}, 104019 (2010).

\bibitem{b3} M. Sasaki, T. Suyama, T. Tanaka, S. Yokoyama, Class. Quantum Gravity {\bf 35}, 063001 (2018)

\bibitem{b4} J. C. Niemeyer, K. Jedamzik, Phys. Rev. {\bf D 59}, 124013 (1999)

\bibitem{b5} H. Lü, A. Perkins, C. N. Pope, and K. S. Stelle, Phys. Rev. Lett., {\bf 114}, 171601 (2015).

\bibitem{b6} B. P. Abbott {\it et al.} (collaboration LIGO Scientific, Virgo), Phys. Rev. Lett. {\bf 116}, 061102 (2016). 

\bibitem{b7} R. Abbott {\it et al.} (collaboration LIGO Scientific, Virgo), Phys. Rev. Lett. {\bf 125}, 101102 (2020).

\bibitem{b8} R. Abbott {\it et al.} (collaboration LIGO Scientific, Virgo), Phys. Rev. {\bf D 102}, 043015 (2020). 

\bibitem{b9} K. Akiyama {\it et al.}, Astrophys. J. {\bf 875}, L1 (2019). 

\bibitem{b10} J. Liu {\it et al.}, Nature {\bf 575}, 618 (2019). 

\bibitem{a4} S. W. Hawking, Phys. Rev. {\bf D 14}, 2460 (1976).

\bibitem{a5} W. G. Unruh, and R. Schützhold, Phys. Rev. {\bf D 68}, 024008 (2003).

\bibitem{a6} S. W. Hawking, Phys. Rev. Lett. {\bf 15}, 689 (1965).

\bibitem{a7} G. W. Gibbons, and S. W. Hawking, Commun. Math. Phys. {\bf 66}, 291 (1979).

\bibitem{a9} M. C. Begelman, R. D. Blandford, and M. J. Rees, Nature {\bf 287}, 307 (1980).

\bibitem{a10} A. Ori, Phys. Rev. Lett. {\bf 67}, 789 (1991).

\bibitem{a11} F. Echeverria, Phys. Rev. {\bf D 40}, 3194 (1989).

\bibitem{a15} P. Hajicek, Phys. Rev. {\bf D 36}, 1065 (1987).

\bibitem{a16} N. Popławski, Astrophys. J. {\bf 832}, 96 (2016).

\bibitem{a17} D. V. Fursaev, Phys. Rev. {\bf D 51}, R5352 (1995).

\bibitem{a18} R. Konoplya, and A. Zhidenko, Phys. Lett. {\bf B 756}, 350 (2016).

\bibitem{a19} S. W. Hawking, Nucl. Phys. {\bf B 239}, 257 (1984).

\bibitem{a20} S. W. Hawking, Phys. Rev. {\bf D 13}, 191 (1976).

\bibitem{a21} S. W. Hawking, and D. N. Page, Commun. Math. Phys. {\bf 87}, 577 (1983).

\bibitem{a22} P. C. Davies, Rep. Prog. Phys.{\bf 41}, 1313 (1978).

\bibitem{a23} J. D. Bekenstein, Phys. Rev. {\bf D 12}, 3077 (1975).

\bibitem{a24} S. W. Hawking, Commun. Math. Phys. {\bf 25}, 167 (1972).

\bibitem{a25} D. L. Wiltshire, Phys. Rev. {\bf D 38}, 2445 (1988).

\bibitem{c1} A. De Felice, S. Tsujikawa, Living Rev. Relativ. {\bf 13}, 3 (2010).

\bibitem{c2} Y.-S. Song, H. Peiris, H. Wayne, Phys. Rev. {\bf D 76}, 063517 (2007)

\bibitem{c3} S. Capozziello, V.F. Cardone, A. Troisi, Phys. Rev. {\bf D 71}, 043503 (2005)

\bibitem{c4} J. C. Fabris, E. L. Junior, and M. E. Rodrigues, Eur. Phys. J. {\bf C 83}, 884 (2023).

\bibitem{c5} V. De Falco, F. Bajardi, R. D’Agostino, M. Benetti, and S. Capozziello, Eur. Phys. J. {\bf C 83}, 456 (2023).

\bibitem{c6} S. E. Jorás, Int. J. Mod. Phys. {\bf A 26}, 3730 (2011)

\bibitem{c7} A. Casado-Turrión, A. Dobado, Á. de la Cruz-Dombriz, Phys. Rev. {\bf D 105}, 084060 (2022)

\bibitem{c8} T. Harko, F. S. N. Lobo, S. I. Nojiri, S. D. Odintsov, Rev. {\bf D 84}, 024020 (2011).

\bibitem{h1} E. T. Akhmedov, T. L. Chau, P. M. Ho, H. Kawai, W. H., Shao, and C. T. Wang, Phys. Rev. {\bf D 109}, 025001 (2024). 

\bibitem{h2} T. McMaken, and A. J. Hamilton, Phys. Rev. {\bf D 107}, 085010 (2023).

\bibitem{h3} R. A. Konoplya, D. Ovchinnikov, and B. Ahmedov, Phys. Rev. {\bf D 108}, 104054 (2023).

\bibitem{h4} R. Bousso, and M. Miyaji, Phys. Rev. {\bf D 109}, 026006 (2024).

\bibitem{h5} Y. Osawa, K. N. Lin, Y. Nambu, M. Hotta, and P. Chen, Phys. Rev. {\bf D 110}, 025023 (2024).

\bibitem{d7} S. Hod, Phys. Rev. {\bf D 80}, 064004 (2009).

\bibitem{d4} T. Regge and J. A. Wheeler, Phys. Rev. {\bf 108}, 1063 (1957).

\bibitem{d5} F. J. Zerilli, Phys. Rev. Lett. {\bf 24}, 737 (1970).

\bibitem{d6} F. J. Zerilli, Phys. Rev. {\bf D 2}, 2141 (1970).

\bibitem{d8} P. A. González, Y. Vásquez, and R. N. Villalobos, Eur. Phys. J. C {\bf 77}, 579 (2017).

\bibitem{d9} R. Konoplya and A. Zhidenko, Phys. Lett. {\bf B 756}, 350 (2016).

\bibitem{d10} J. Maldacena, Int. J. Theor. Phys. {\bf 38}, 1113 (1999).

\bibitem{d11} G. T. Horowitz and V. E. Hubeny, Phys. Rev. {\bf D 62}, 024027 (2000).

\bibitem{d1} R. A. Konoplya and A. Zhidenko, Rev. Mod. Phys. {\bf 83}, 793 (2011).

\bibitem{Amendola:2020qho} L.~Amendola, L.~Giani and G.~Laverda, Phys. Lett. B \textbf{811}, 135923 (2020).

\bibitem{Malik:2024zhb} A.~Malik, A.~Arif and M.~F.~Shamir, Eur. Phys. J. Plus \textbf{139}, 67 (2024).

\bibitem{Ahmed:2024jsh} F.~Ahmed and A.~Bouzenada, Eur. Phys. J. C {\bf 84}, 1271 (2024).

\bibitem{Moreira:2024mkb} A.~R.~P.~Moreira, F.~Ahmed and S.~H.~Dong, Ann. Phys. (NY) \textbf{469}, 169763 (2024).

\bibitem{Ahmed:2024jcb} F.~Ahmed and A.~Guvendi, Chin. J. Phys. \textbf{89}, 69 (2024).

\bibitem{Mustafa:2023mls} G.~Mustafa, Phys. Lett. B \textbf{848}, 138407 (2024).

\bibitem{Malik:2023yaj} A.~Malik, E.~Meer, Z.~Asghar and A.~Ali, Chin. J. Phys. \textbf{86}, 391 (2023).

\bibitem{ref4} T. Q. Do, Eur. Phys. J. C {\bf 81}, 431 (2021).

\bibitem{ref5} T. Q. Do, Eur. Phys. J. C {\bf 82}, 15 (2022).

\bibitem{ref8} M. F. Shamir, M. Ahmad, G. Mustafa, and A. Rashid, Chin. J. Phys. {\bf 81}, 51 (2023).

\bibitem{ref9} A. Jawad, and A. M. Sultan, EPL {\bf 138}, 29001 (2022).

\bibitem{plb} G. Mustafa, Phys. Lett. {\bf B 848}, 138407 (2024).

\bibitem{meer} M. F. Shamir, and E. Meer, Eur. Phys. J. C {\bf 83}, 49 (2023).

\bibitem{AM1} A. Malik, A. Hussain, M. Ahmad, and M. F. Shamir, Chin. J. Phys. 91, 560 (2024).

\bibitem{AM2} A. Malik, A. Hussain, M. Ahmad, and M. F. Shamir, Eur. Phys. J. Plus {\bf 139}, 535 (2024).

\bibitem{AM3} G. Mustafa, F. Javed, S. K. Maurya, and A. Errehymy, Ann. Phys. (Berlin) 536, 2400155 (2024).

\bibitem{MFS2} A. Malik, A. Arif, M. F. Shamir, Eur. Phys. J. Plus {\bf 139}, 67 (2024).

\bibitem{ijtp} A. Malik, A. Arif, and M. F. Shamir, Int. J. Theor. Phys. {\bf 62}, 243 (2023). 

\bibitem{aop} F. Ahmed, J. C. R. de Souza, and A. F. Santos, Ann. Phys. (N Y) {\bf 461}, 169578 (2024).

\bibitem{EPJP} F. Ahmed, J. C. R. de Souza, and A. F. Santos, Eur. Phys. J Plus {\bf 139}, 419 (2024).

\bibitem{NA} F. Ahmed, New Astron. {\bf 111}, 102245 (2024).

\bibitem{NPB} F. Ahmed, J. C. R. de Souza, and A. F. Santos, Nucl. Phys. {\bf B 1004}, 116573 (2024).

\bibitem{EPJC} J. C. R. de Souza, A. F. Santos, and F. Ahmed, Eur. Phys. J. C {\bf 84}, 559 (2024).

\bibitem{EPJC3} A. R. P. Moreira, S.-H. Dong, and F. Ahmed, Eur. Phys. J. C {\bf 84}, 913 (2024).

\bibitem{EPJC4} F. Ahmed, J. C. R. de Souza, and A. F. Santos, Eur. Phys. J. C {\bf 84}, 968 (2024).

\bibitem{JCAP} F. Ahmed, J. C. R. de Souza, and A. F. Santos, JCAP 10 ({\bf 2024}) 015.

\bibitem{arxiv} F. Ahmed and A. Bouzenada, arXiv: 2411.00896 [gr-qc] \url{https://doi.org/10.48550/arXiv.2411.00896}.

\bibitem{JPSL} J. P. S. Lemos, Phys. Lett. {\bf B 353}, 46 (1995).

\bibitem{LL} L. D. Landau, and E. M. Lifshitz, {\it The Classical Theory of Fields}, Pergamon (1975).

\bibitem{AA1} M. Bouhmadi-Lopez, S. Brahma, C.-Y. Chen, P. Chen, and D.-H. Yeom, JCAP {\bf 07}, 066 (2020).

\bibitem{Hawking:1974rv} S.~W.~Hawking, Nature \textbf{248}, 30 (1974).

\bibitem{Gibbons:1977mu} G.~W.~Gibbons and S.~W.~Hawking, Phys. Rev. D \textbf{15}, 2738 (1977).

\bibitem{Hawking:1982dh} S.~W.~Hawking and D.~N.~Page, Commun. Math. Phys. \textbf{87}, 577 (1983).

\bibitem{Parikh:1999mf} M.~K.~Parikh and F.~Wilczek, Phys. Rev. Lett. \textbf{85}, 5042 (2000).

\bibitem{Srednicki:1993im} M.~Srednicki, Phys. Rev. Lett. \textbf{71}, 666 (1993).

\bibitem{Hawking:1976de} S.~W.~Hawking, Phys. Rev. D \textbf{13}, 191 (1976).

\bibitem{Callan:1992rs} C.~G.~Callan, Jr., S.~B.~Giddings, J.~A.~Harvey and A.~Strominger, Phys. Rev. D \textbf{45}, R1005 (1992).

\bibitem{Kanzi:2019gtu} S.~Kanzi and \.I.~Sakall\i{}, Nucl. Phys. B \textbf{946}, 114703 (2019)

\bibitem{Kanzi:2021cbg} S.~Kanzi and \.I.~Sakall\i{}, Eur. Phys. J. C \textbf{81}, 501 (2021).

\bibitem{Ovgun:2019ygw} A.~\"Ovg\"un and \.I.~Sakall\i{}, Ann. Phys. (NY) \textbf{413}, 168071 (2020).

\bibitem{Sakalli:2018nug} \.I.~Sakall\i{}, K.~Jusufi and A.~\"Ovg\"un, Gen. Rel. Grav. \textbf{50}, 125 (2018).

\bibitem{Al-Badawi:2020htj} A.~Al-Badawi, S.~Kanzi and \.I.~Sakall\i{}, Eur. Phys. J. Plus \textbf{135}, 219 (2020).

\bibitem{Sakalli:2023pgn} \.I.~Sakall\i{} and E.~Y\"or\"uk, Phys. Scr. \textbf{98}, 125307 (2023).

\bibitem{Sucu:2023moz} E.~Sucu and \.I.~Sakall\i{}, Phys. Scr. \textbf{98}, 105201 (2023).

\bibitem{Kanzi:2023itu} S.~Kanzi, \.I.~Sakall\i{} and B.~Pourhassan, Symmetry \textbf{15}, 873 (2023).

\bibitem{Maggiore:1993rv} M.~Maggiore, Phys. Lett. B \textbf{304}, 65 (1993).

\bibitem{Scardigli:1999jh} F.~Scardigli, Phys. Lett. B \textbf{452}, 39 (1999).

\bibitem{Konishi:1989wk} K.~Konishi, G.~Paffuti and P.~Provero, Phys. Lett. B \textbf{234}, 276 (1990).

\bibitem{Sakalli:2022swm} \.I.~Sakall\i{} and S.~Kanzi, Ann. Phys. (NY) \textbf{439}, 168803 (2022).

\bibitem{Sakalli:2016mnk} I.~Sakalli, A.~\"Ovg\"un and K.~Jusufi, Astrophys. Space Sci. \textbf{361}, 330 (2016).

\bibitem{Wald:1979kp} R.~M.~Wald, Commun. Math. Phys. \textbf{70}, 221 (1979).

\bibitem{Visser:2021ucg} M.~Visser, J High Energy Phys. \textbf{2022}, 129 (2022).

\bibitem{Akhmedova:2008dz} V.~Akhmedova, T.~Pilling, A.~de Gill and D.~Singleton, Phys. Lett. B \textbf{666}, 269 (2008).

\bibitem{Akhmedova:2008au} V.~Akhmedova, T.~Pilling, A.~de Gill and D.~Singleton, Phys. Lett. B \textbf{673}, 227 (2009).

\bibitem{Wald:1984rg} R.~M.~Wald, \textit{General Relativity} (Chicago University Press, Chicago, 1984).

\bibitem{Aydemir:2020pao} U.~Aydemir, B.~Holdom and J.~Ren, Phys. Rev. D \textbf{102}, 083009 (2020).

\bibitem{Fernandes:2021dsb} P.~G.~S.~Fernandes, P.~Carrilho, T.~Clifton and D.~J.~Mulryne, Phys. Rev. D \textbf{104}, 044029 (2021).


\bibitem{Medved:2004yu} A.~J.~M.~Medved and E.~C.~Vagenas, Phys. Rev. D \textbf{70}, 124021 (2004).

\bibitem{Khosropour:2024boy} B.~Khosropour, Mod. Phys. Lett. A \textbf{39}, 2450120 (2024)

\bibitem{Chen:2024mlr} H.~Chen, S.~H.~Dong, E.~Maghsoodi, S.~Hassanabadi, J.~K\v{r}i\v{z}, S.~Zare and H.~Hassanabadi, Eur. Phys. J. Plus \textbf{139}, 759 (2024).

\bibitem{Sucu:2024gtr} E.~Sucu, arXiv:2404.13450 [gr-qc] \url{https://doi.org/10.48550/arXiv.2404.13450}.

\bibitem{Sachdev:2015efa} S.~Sachdev, Phys. Rev. X \textbf{5}, 041025 (2015).

\end{thebibliography}
\end{document}